\documentclass[aps,prc,amsfonts,preprintnumbers,superscriptaddress,nofootinbib]{revtex4}
\usepackage{graphics}
\usepackage{amsmath}
\usepackage{hyperref}
\usepackage{amssymb}
\usepackage{psfrag}
\usepackage{epsfig}
\usepackage{epsf}
\usepackage{float}
\usepackage{slashed}
\usepackage{dsfont}
\usepackage{bbold}
\usepackage{color}
\allowdisplaybreaks

\newcommand{\fet}[1]{\mbox{\boldmath $#1$}}

\newcommand{\beq}{\begin{equation}}
\newcommand{\eeq}{\end{equation}}
\newcommand{\beqa}{\begin{eqnarray}}
\newcommand{\eeqa}{\end{eqnarray}}
\newcommand{\nn}{\nonumber \\ }

\begin{document}

\title{Towards consistent nuclear interactions from chiral
  Lagrangians II: \\ Symmetry preserving regularization}

\author{H.~Krebs}
\email[]{Email: hermann.krebs@rub.de}
\affiliation{Institut f\"ur Theoretische Physik II, Ruhr-Universit\"at Bochum,
  D-44780 Bochum, Germany}
\author{E.~Epelbaum}
\email[]{Email: evgeny.epelbaum@rub.de}
\affiliation{Institut f\"ur Theoretische Physik II, Ruhr-Universit\"at Bochum,
  D-44780 Bochum, Germany}

\date{\today}

\begin{abstract}
Low-energy nuclear structure and reactions can be described in a
systematically improvable way using the framework of chiral effective
field theory. This requires 
solving the quantum mechanical many-body problem with regularized
nuclear forces and current operators, derived from the most general
effective chiral Lagrangian. To maintain the chiral and gauge
symmetries, a symmetry preserving cutoff regularization has to be employed when
deriving nuclear potentials. Here, we discuss various 
regularization techniques and show how this task
can be accomplished by regularizing the pion field in the effective chiral
Lagrangian using the gradient flow method. The actual derivation of
the nuclear forces and currents from the regularized effective Lagrangian can be carried
out utilizing the novel path-integral approach introduced in our earlier paper. 
\end{abstract}


\maketitle

\vspace{-0.2cm}

\section{Introduction}
\def\theequation{\arabic{section}.\arabic{equation}}
\label{sec:Intro}

Dimensional regularization (DR) is a
commonly used method to treat ultraviolet divergent loop integrals in quantum
field theory.
Its usefulness rests on its computational efficiency and the fact that
this method
preserves the relevant symmetries such as the chiral and gauge
ones. However, while being highly convenient in the realm of
perturbation theory, DR can usually not be
employed in nonperturbative calculations, especially if results are
only available numerically as it is, e.g., the case in
applications of chiral effective field theory (EFT)
to nuclear systems. Here, observables are calculated by 
a numerical solution of the $A$-body Schr\"odinger equation using (highly
singular) potentials derived from the effective chiral Lagrangian, see
Refs.~\cite{Epelbaum:2008ga,Machleidt:2011zz} for review articles. The (unphysical) singular short-distance
behavior of such potentials reflects their effective nature, and it is
usually dealt with by introducing some kind of cutoff regulator in the
Schr\"odinger equation. Renormalization then ensures that the calculated
low-energy observables are insensitive to regularization details at
the accuracy level of the EFT expansion \cite{Gasparyan:2021edy,Gasparyan:2023rtj}. 

In Refs.~\cite{Epelbaum:2019kcf,Krebs:2019uvm}, we pointed out that
the commonly used approach to regularize pion-exchange potentials by
multiplying the expressions, derived in chiral EFT using
momentum-independent  regularization schemes
such as DR, with some cutoff functions is inconsistent as it violates
chiral symmetry when applied to three- and
more-nucleon systems and/or processes involving external probes. The
inconsistency originates from mixing the two different regularization
schemes. It can be avoided by re-deriving nuclear potentials
using cutoff regularization instead of DR.
The purpose of this paper is to discuss in detail symmetry-preserving
regularization of the effective chiral Lagrangian, which is suitable for
the derivation of regularized nuclear interactions from chiral EFT.
Specifically, we focus here on cutoff regularization
of the pion and/or pion-nucleon Lagrangians $\mathcal{L}_{\pi}$ and
$\mathcal{L}_{\pi N}$, respectively, and we seek for a scheme that (i) preserves
the chiral and gauge symmetries, (ii) reduces to the Gaussian-type regulator of Ref.~\cite{Reinert:2017usi}
when applied to the one-pion exchange two-nucleon potential and (iii)
results in finite pion-exchange contributions to the nuclear forces 
and current operators (at least if used in combination with DR or
$\zeta$-function regularization), which are
non-singular at short distances (i.e., sufficiently regularized). 
We emphasize that at the accuracy level of chiral EFT we aim for, regularization of
few-nucleon contact interactions does not require additional efforts
to maintain the chiral symmetry. The regularized chiral
Lagrangian constructed in this paper can be employed to derive
\emph{consistent} nuclear interactions using the path-integral
approach introduced in Ref.~\cite{Krebs:2023ljo}.

Our paper is organized as follows. First, in sec.~\ref{sec:HDRNaive} we briefly remind the
reader of the main idea behind the higher-derivative method introduced
by Slavnov to regularize the non-linear $\sigma$-model
\cite{Slavnov:1971aw} before DR was invented, see also
Refs.~\cite{Djukanovic:2004px,Long:2016vnq} for a related work. 
We also show in this section that a naive application of this method to the problem at
hand fails because of deregularization of multi-pion
interactions. This issue can be mitigated by choosing all 
cutoff-dependent terms in the effective Lagrangian to be
proportional to the equation of motion (EOM), as done in the original work by Slavnov
\cite{Slavnov:1971aw}.  In sec.~\ref{sec:HDR}, we consider such an ansatz for the
regularized pion Lagrangian and show that it indeed leads to
satisfactory results as far as nuclear forces are concerned. However, when applied to
the electromagnetic current operator, this scheme leads to certain
badly divergent loop integrals with exponentially growing integrands,
signalling that the considered ansatz is still insufficient
for our purposes. An alternative approach to regularize the
pion field using the gradient flow method has been suggested by 
Kaplan \cite{Kaplan} and  is considered in sec.~\ref{sec:GradientFlow}. We work out the explicit
form of the regularized effective Lagrangian and show that this scheme
yields finite\footnote{Certain loop integrals that involve
  only pion propagators
  remain unregularized in the gradient flow method, but they can be
  dealt with by using
  an additional symmetry-preserving regularization such as, e.g., DR.} and
regularized expressions for the nuclear forces and
currents, thereby fulfilling all demanded criteria. 
The main results of
our paper are summarized in sec.~\ref{sec:summary}, while appendices
\ref{Derivation_of_4NF},
\ref{Derivation_of_EOM} and \ref{AppRegularizedLagrangian} 
provide details on the derivation of the four-nucleon force considered
in the main text, the equation of motion (EOM) for pion fields and the
regularized Lagrangian using the ansatz of
sec.~\ref{sec:HDR}. Finally, appendix 
\ref{formal_proofs_gradient_flow} describes the 
proof that the chiral transformation properties of the generalized pion field
in the gradient flow method do not depend on the flow parameter.


\section{Higher derivative regularization}
\def\theequation{\arabic{section}.\arabic{equation}}
\label{sec:HDRNaive}

We start with the effective chiral Lagrangian, which can be
conveniently written in terms of the $\rm SU(2)$ matrix
$U$ that collects the pion fields and transforms with respect to the
chiral  $\rm SU(2)_L \times SU(2)_R$ rotations according to $U \to R U
L^\dagger$. Here and in what follows, we use the standard notation by
introducing the building
blocks~\cite{Bernard:1995dp} 
\beqa
u&=&\sqrt{U},\quad\quad
u_\mu\, =\, iu^\dagger \nabla_\mu U u^\dagger, \quad\quad
\nabla_\mu U\,=\,\partial_\mu U -ir_\mu U + i U l_\mu,\quad\quad
D_\mu\,=\,\partial_\mu+\Gamma_\mu,\nn
\Gamma_\mu&=&\frac{1}{2}\big[u^\dagger,\partial_\mu
u\big]-\frac{i}{2} u^\dagger r_\mu u - \frac{i}{2} u\, l_\mu
u^\dagger,\quad\quad
\chi_{\pm}\,=\,u^\dagger \chi\, u^\dagger \pm u \,\chi^\dagger u, \quad\quad
\chi\,=\,2 B(s+i p),
\label{ChiralBuildingBlocks}
\eeqa
where $\Gamma_\mu$ is the chiral connection,  $l_\mu$, $r_\mu$,
$s$ and $p$ refer to left-handed, right-handed, scalar and
pseudoscalar external sources, respectively, while $B$ is a low-energy
constant. Throughout this paper, we neglect isospin-breaking corrections
and set the external scalar source $s$ to a constant value equal to the
average mass of the up- and down-quarks, $s=m_q$. 

To derive nuclear potentials using the path-integral
method of Ref.~\cite{Krebs:2023ljo}, we have
to switch to Euclidean space. The lowest-order pion Lagrangian has the well-known form
\beq
\label{LagrLO}
{\cal L}_\pi^{\rm E\, (2)} = \frac{F^2}{4}{\rm Tr}\big[u_\mu u_\mu -
\chi_+\big]
\, =\,  \frac{F^2}{4}{\rm Tr}\big[\big(\nabla_\mu
U\big)^\dagger \nabla_\mu U- U^\dagger\chi-\chi^\dagger U\big]\,,
\eeq
where $F$ is the pion decay constant in the chiral limit. The
superscripts $\rm E$ and $(2)$ signify that the Lagrangian is written in
Euclidean space and that it involves terms with two derivatives or a
single insertion of the quark mass $m_q$, respectively.  

Our goal is to modify the effective Lagrangian by introducing a
momentum-space cutoff in a symmetry-preserving manner to ensure 
that all pion-exchange 
contributions to the nuclear potentials and currents get
sufficiently regularized with respect to the external momenta of the
nucleons.
As already pointed out in our earlier papers
\cite{Epelbaum:2019kcf, Krebs:2019uvm,Krebs:2020plh,Epelbaum:2022cyo}, the higher derivative method of Ref.~\cite{Slavnov:1971aw} provides a
promising approach to achieve this goal. 
The main idea behind this method is to modify
the Lagrangian by including certain higher-order terms, chosen to
improve the ultraviolet behavior of loop integrals that appear in
the calculation of the scattering amplitude. 
Below, we demonstrate this idea by considering an explicit  example,
and we also point out some subtleties that occur in a
naive application of this scheme.
For the sake of simplicity, we switch off all external sources and consider the
chiral limit of $m_q = 0$ throughout
this section.

Inspired by
the original work by Slavnov \cite{Slavnov:1971aw}, we intend to improve the
ultraviolet behavior of the pion propagator by modifying the
Lagrangian via:
\beqa
{\cal L}_{\pi}^{\rm E\, (2)}&=&\frac{F^2}{4}{\rm Tr}\Big[\partial_\mu
U^\dagger\partial_\mu U\Big]\, \,\longrightarrow \,\,  {\cal L}_{\pi,\Lambda}^{\rm E\, (2)}\,=\,\frac{F^2}{4}{\rm
  Tr}\bigg[\partial_\mu U^\dagger e^{- \partial^2/\Lambda^2} \partial_\mu U\bigg],\label{Lpi2LambdaTrial}
\eeqa
where $\Lambda$ is a momentum cutoff. The choice of the modification
is obviously not unique: while Ref.~\cite{Slavnov:1971aw}
included terms polynomial in $\partial^2$, the modification in
Eq.~(\ref{Lpi2LambdaTrial})
is chosen to match the regularized one-pion exchange
two-nucleon potential of Refs.~\cite{Reinert:2017usi,Reinert:2020mcu}. However, we
will show below that the ansatz in Eq.~(\ref{Lpi2LambdaTrial}) does not
provide a sufficient regularization for certain type of contributions
to the nuclear forces and is, therefore, not sufficient for our
purposes.   

\begin{figure}[tb]
  \begin{center} 
\includegraphics[width=0.5\textwidth,keepaspectratio,angle=0,clip]{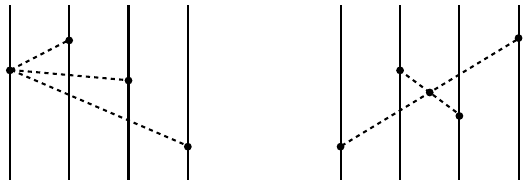}
\caption{
Representation invariant class of diagrams that contribute to the
four-nucleon force at N$^3$LO. Solid and dashed lines refer to nucleons and
pions, respectively. Solid dots denote the lowest-order vertices from
the effective chiral Lagrangian. 
  \label{fig:1} 
 }
  \end{center}
\end{figure}

To proceed with the calculation, we employ the so-called
$\sigma$-gauge parametrization of the matric $U$ in terms of the pion field
$\fet \pi$,  
\beqa
\label{SigmaModG}
U&=&\sqrt{1-\frac{\fet{\pi}^2}{F^2}}+i\frac{\fet{\tau}\cdot\fet{\pi}}{F}\,,
\eeqa
where $\fet \tau$ refer to the isospin Pauli matrices. 
Then, the regularized Lagrangian of
Eq.~(\ref{Lpi2LambdaTrial}), expanded in powers of the pion field,
takes the form:
\beqa
 {\cal
   L}_{\pi,\Lambda}^{\rm E\, (2)}&=&- \frac{1}{2}\fet{\pi}\cdot \partial^2
 \, e^{-\partial^2/\Lambda^2}\fet{\pi}+\frac{1}{2
   F^2} \fet{\pi}\cdot\partial_\mu\fet{\pi} \, e^{-\partial^2/\Lambda^2}
 \fet{\pi}\cdot\partial_\mu\fet{\pi}+{\cal O}(\pi^6).\label{Lpi2ChiralLimitExpandedSigmaGauge}
 \eeqa
The first term can be rewritten in terms of the inverse pion
propagator. As desired, the exponentially growing (in momentum space) factor in
this term improves the ultraviolet behavior of the pion
propagator in Euclidean space, which takes the form (in the chiral limit)
\beq
\Delta^{\rm E} (x_i - x_j )  \, = \,
\int \frac{d^4q}{(2\pi)^4} e^{i q\cdot (x_i-x_j)} \,
\frac{e^{-(q_0^2+\vec{q}^{\, 2})/\Lambda^2}}{q_0^2+\vec{q}^{\, 2}}.
\label{PionPropCL}
\eeq
However, from
Eq.~(\ref{Lpi2ChiralLimitExpandedSigmaGauge}) we see that
also interactions involving four and more pion fields acquire exponential factors that grow with increasing
momenta. Such factors are problematic from the regularization point of
view as can already be seen from tree-level diagrams involving the four-pion
vertex. In Fig.~\ref{fig:1}, we show selected contributions to the
four-nucleon force (4NF) at fourth order (N$^3$LO) in chiral EFT.
A complete set of diagrams and the corresponding unregularized expressions for the
4NF, calculated using the method of unitary transformation \cite{Epelbaum:1998ka}, can be
found in Refs.~\cite{Epelbaum:2005bjv,Epelbaum:2007us}. Notice that
while the contribution of each of the shown Feynman diagrams depends on the
parametrization of the SU(2) matrix $U$, their sum
is representation invariant. Since the first diagram does not involve any
vertices from ${\cal L}_{\pi,\Lambda}^{\rm E\, (2)}$, its contribution
can be read off from the corresponding unregularized result by simply
replacing all pion propagators with the regularized ones given in Eq.~(\ref{PionPropCL}). More
interesting is the 4NF contribution stemming from the second diagram in
Fig.~\ref{fig:1}. In Ref.~\cite{Krebs:2023ljo}, we have introduced a general method
to derive nuclear forces and currents from regularized effective
Lagrangians using the path-integral formalism. However, since the
diagram we are interested in does not involve reducible pieces, the
expression for the 4NF coincides with the one for the scattering amplitude and
can be easily obtained using the standard Feynman diagram technique.
A straightforward calculation using the lowest-order pion-nucleon
Lagrangian, along with the regularized Lagrangian ${\cal
  L}_{\pi,\Lambda}^{\rm E\, (2)}$ of
Eq.~(\ref{Lpi2ChiralLimitExpandedSigmaGauge}) yields the following result for the second
graph in Fig.~\ref{fig:1} (in the chiral
limit)\footnote{For illustrative purposes, we also derive in
  appendix \ref{Derivation_of_4NF} the resulting 4NF using the
  path-integral approach of Ref.~\cite{Krebs:2023ljo}.}:
\beqa
V^{4N}_{\Lambda} &=& \frac{g^4}{128 F^6}\fet \tau_1 \cdot \fet \tau_2 \fet \tau_3 \cdot \fet
\tau_4 \,
\frac{\vec \sigma_1 \cdot \vec q_1 \vec \sigma_2 \cdot \vec q_2 \vec
  \sigma_3 \cdot \vec q_3 \vec \sigma_4 \cdot \vec q_4}{\vec q_1^{\,
    2} \, \vec q_2^{\,
    2} \, \vec q_3^{\,
    2} \, \vec q_4^{\,
    2}} \, \vec q_{12}^{\, 2}\, e^{-\frac{\vec q_1^{\,
    2}}{\Lambda^2}}\, e^{-\frac{\vec q_2^{\,
    2}}{\Lambda^2}}\, e^{-\frac{\vec q_3^{\,
    2}}{\Lambda^2}}\, e^{-\frac{\vec q_4^{\,
    2}}{\Lambda^2}} \, 
e^{\frac{\vec q_{12}^{\, 2}}{\Lambda^2}} \, +\,  23 \text{ perm.} \nn[3pt]
&=& \frac{g^4}{128 F^6}\fet \tau_1 \cdot \fet \tau_2 \fet \tau_3 \cdot \fet
\tau_4 \,
\frac{\vec \sigma_1 \cdot \vec q_1 \vec \sigma_2 \cdot \vec q_2 \vec
  \sigma_3 \cdot \vec q_3 \vec \sigma_4 \cdot \vec q_4}{\vec q_1^{\,
    2} \, \vec q_2^{\,
    2} \, \vec q_3^{\,
    2} \, \vec q_4^{\,
    2}} \,  \vec q_{12}^{\, 2}\,
e^{-\frac{\vec q_{13}^{\, 2}}{\Lambda^2}} 
e^{-\frac{\vec q_{14}^{\, 2}}{\Lambda^2}} 
\, +\,  23 \text{ perm.} ,
\label{4NFNaive}
\eeqa
where we have dropped the prefactor 
$(2 \pi)^3 \delta (\vec q_1 + \vec q_2 + \vec q_3 + \vec q_4 )$
following the standard convention for nuclear potentials. Here,
$\vec \sigma_i$ and 
$\vec q_i = \vec p_i^{\, \prime } - \vec p_i$ denote the Pauli spin
matrices and the momentum transfer of
nucleon $i$, respectively,
while $\vec q_{ij} \equiv \vec q_i + \vec q_j$. Further, $g$ denotes the
nucleon axial vector coupling in the chiral limit. To obtain the last equality in Eq.~(\ref{4NFNaive}),
we have used the obvious relationship $\vec q_1 + \vec q_2 + \vec q_3 + \vec
q_4 = 0$. When removing the regulator by taking the limit $\Lambda \to
\infty$, the above expression reduces to the chiral-limit version of Eq.~(3.44) of
Ref.~\cite{Epelbaum:2007us} if one sets $\alpha = 0$ to comply with 
the $\sigma$-gauge parametrization in Eq.~(\ref{SigmaModG}). 
From Eq.~(\ref{4NFNaive}), we see that the resulting
four-nucleon force is not sufficiently regularized. To ensure a convergent
behavior when solving the nuclear Schr\"odinger
equation, the 4NF should be
regularized in at least three independent combinations $(\sum_i
\alpha_i \vec
q_i)^2$,
while the obtained expression is only regularized with respect to the
momenta $\vec q_{13}^{\, 2}$ and  $\vec q_{14}^{\, 2}$.
Notice that the problematic deregularization is not actually caused by the
appearance of the exponentially increasing factor in the first
equality of Eq.~(\ref{4NFNaive}), which originates from the last term
in Eq.~(\ref{Lpi2ChiralLimitExpandedSigmaGauge}), but rather 
by its dependence on the linear combination $\vec{q}_1 + \vec{q}_2$. 
Were it dependent on the
momentum transfer of a single nucleon only, then the 4NF
would  possess an exponentially
decreasing behavior in  three momenta, thus being sufficiently regularized.
As already pointed out in the Introduction, this issue can be
mitigated using a more sophisticated ansatz for the additional cutoff-dependent terms in the Lagrangian
by requiring them to be proportional to the EOM.
While not explicitly emphasized, a closer look
at the original paper by Slavnov~\cite{Slavnov:1971aw} reveals that
all additional cutoff-dependent terms in the Lagrangian were chosen to
be proportional to the EOM. In the next section we consider a more elaborate
ansatz for the regularized Lagrangian along this line.


\section{Higher derivative regularization using terms proportional to the
  EOM}
\def\theequation{\arabic{section}.\arabic{equation}}
\label{sec:HDR}

We start with the lowest-order pion Lagrangian in
Eq.~(\ref{LagrLO}), which leads to the equation of motion 
given by\footnote{Note that the ${\rm EOM}$ in
  Eq.~(\ref{EOMDefinition}) does not include higher-order pionic
  interactions and terms involving the nucleon fields. It is not
  necessary to take into account such terms for our purpose of introducing a
  regulator.}
\beqa
{\rm EOM}&=&0, \quad \quad {\rm EOM} :=\,\big[D_\mu,
u_\mu\big]+\frac{i}{2}\chi_{-}-\frac{i}{4}{\rm Tr}\,\chi_{-}\,,\label{EOMDefinition}
\eeqa
as derived in Appendix \ref{Derivation_of_EOM}
and make an ansatz for the regularized version of the Lagrangian in the form
\beqa
{\cal L}_{\pi,\Lambda}^{\rm E \, (2)}&=&{\cal L}_\pi^{\rm E \,  (2)} - \frac{F^2}{4}{\rm
  Tr}\left[{\rm EOM}\frac{1-\exp\left(\frac{-{\rm ad}_{D_\mu}{\rm
          ad}_{D_\mu}+\frac{1}{2}\chi_+}{\Lambda^2}\right)}{-{\rm
      ad}_{D_\mu}{\rm ad}_{D_\mu}+\frac{1}{2}\chi_+}{\rm EOM}\right],\label{Lpi2Reg}
\eeqa
where the additional term is constructed to be chirally invariant and
is chosen in
such a way that the pion propagator acquires a Gaussian regulator. 
Here and in what follows, we use the notation ${\rm ad}_A B\,\equiv\,[A,B ]$. 
In the $\Lambda\to \infty$ limit, the second term in Eq.~(\ref{Lpi2Reg})
vanishes, and the Lagrangian reduces to its original non-regularized
form.
In Appendix~\ref{AppRegularizedLagrangian} we show that
the regularized Lagrangian ${\cal L}_{\pi,\Lambda}^{\rm E \, (2)}$, expanded in
powers of the pion field, takes the form
\beqa
\label{regularizedLpi2Expanded}
{\cal L}_{\pi,\Lambda}^{\rm E \, (2)}&=&-F^2
M^2\, +\, \frac{1}{2}\fet{\pi}\cdot\big(-\partial^2+M^2\big) e^{\frac{-\partial^2
  + M^2}{\Lambda^2}}
\fet{\pi}
\, +\, \frac{\fet{\pi}^2}{8F^2}\Big[6\,\partial_\mu\fet{\pi}\cdot\partial_\mu\fet{\pi}+3\,\fet{\pi}\cdot\partial^2\fet{\pi}+\fet{\pi}\cdot(-\partial^2+M^2)\fet{\pi}\Big]\nn[3pt]
&&-\frac{\alpha}{F^2}\fet{\pi}^2\fet{\pi}\cdot\big(-\partial^2+M^2\big)e^{\frac{-\partial^2+M^2}{\Lambda^2}}
\fet{\pi}
\, -\,
\frac{1}{F^2}\bigg(\partial_\mu\fet{\pi}\cdot\partial_\mu\fet{\pi}+\frac{1}{2}\fet{\pi}\cdot\partial^2\fet{\pi}\bigg)\fet{\pi}\cdot
e^{\frac{-\partial^2+M^2}{\Lambda^2}}\fet{\pi}\nn[3pt]
&& + \frac{1}{4 F^2\Lambda^2}\int_0^1
d
s\bigg\{M^2 \fet{\pi}^2\bigg[\bigg(1-
e^{(1-s)\frac{-\partial^2+M^2}{\Lambda^2}}\bigg)\fet{\pi}\bigg]
\cdot\bigg[\big(-\partial^2+M^2\big) \, e^{s\frac{-\partial^2+M^2}{\Lambda^2}}\fet{\pi}\bigg]\\[3pt]
&& \hspace{2.5cm}+ \big(\fet{\pi}\times\partial_\mu\fet{\pi}\big)\cdot\bigg[\bigg(e^{s\frac{-\partial^2+M^2}{\Lambda^2}}\big(-\partial^2+M^2\big)\fet{\pi}\bigg)\times
\overleftrightarrow{\partial}_{\!\!\mu}
\bigg(1-e^{(1-s)\frac{-\partial^2+M^2}{\Lambda^2}}\bigg)\fet{\pi}
\bigg]\bigg\} \, +\,   {\cal O}(\pi^6),
\nonumber
\eeqa
where the external pseudoscalar and electroweak sources have been
switched off while the scalar source is set to a constant value equal
to the light-quark mass $m_q$. In
Eq.~(\ref{regularizedLpi2Expanded}), we use the notation
\beqa
A\cdot\big[B\times\overleftrightarrow{\partial}_{\!\!\mu} C\big]&\equiv&A\cdot\big[B\times\partial_\mu
C\big]-A\cdot\big[\big(\partial_\mu B\big)\times C\big].
\eeqa
Furthermore, $M$ denotes the pion mass at lowest order in the chiral
EFT expansion, while the arbitrary constant $\alpha$ reflects the
freedom in parametrizing the matrix $U$ in terms of the pion field:
\beqa
U&=&1+\frac{i}{F}{\boldsymbol\tau}\cdot{\boldsymbol\pi}\bigg(1-\alpha\frac{\fet
  \pi^2}{F^2}\bigg)-\frac{\fet \pi^2}{2
F^2}\bigg[1+\bigg(\frac{1}{4}-2\alpha\bigg)\frac{\fet \pi^2}{F^2}\bigg]+{\cal
O}(\fet \pi^5)\,.\label{U_expressed_by_pi}
\eeqa
Clearly, observable quantities do not depend on the parametrization of $U$.
As one can see from Eq.~(\ref{regularizedLpi2Expanded}), all
exponential operators act, in contrast to the ansatz considered in the last
section and leading to Eq.~(\ref{Lpi2ChiralLimitExpandedSigmaGauge}), only on a single pion field. For this reason,
all exponentially increasing factors that appear in the four-pion vertex
involve single-pion momenta and are angle-independent. 

It is instructive to look at the regularized 4NF contributions
stemming from the Feynman diagrams shown in Fig.~\ref{fig:1}. 
Using the regularized Lagrangian in
Eq.~(\ref{Lpi2Reg}) in combination with the lowest-order pion-nucleon
Lagrangian
\beq
\mathcal{L}_{\pi N}^{\rm E \, (1)} = N^\dagger \bigg[\partial_0 - \frac{g}{2
  F} \vec \sigma \cdot \vec \nabla \fet \pi \cdot \fet \tau
+\frac{i}{4 F^2}(\fet\tau\times\fet\pi)\cdot\partial_0\fet\pi
+ \frac{g}{2 F^3}\bigg( 2 \alpha - \frac{1}{2} \bigg) \fet \tau \cdot \fet \pi \,
\fet \pi \vec \sigma \cdot \vec \nabla \fet \pi + \frac{g}{2 F^3} \alpha \fet \pi^2 \,
\vec \sigma \cdot \vec \nabla \fet \pi \cdot \fet \tau \bigg] N +
\ldots\,,
\label{LagrPiNNoReg}
\eeq
where the ellipses refer to terms with at least four pion fields, the resulting
4NF takes the form
\beqa
V^{4N}_{\Lambda} &=& -\frac{g^4}{128 F^6} \fet \tau_1 \cdot \fet \tau_2 \fet
\tau_3 \cdot \fet \tau_4 \frac{\vec \sigma_2 \cdot \vec q_2\, 
  \vec \sigma_3 \cdot \vec q_3\, 
  \vec \sigma_4 \cdot \vec q_4
}{(\vec q_2^{\, 2} + M^2) (\vec q_3^{\, 2} + M^2) (\vec q_4^{\, 2} +
  M^2)}
\Big[ 2 \vec \sigma_1 \cdot \vec q_{12} f_\Lambda^{234} + 
\vec \sigma_1 \cdot \vec q_{1} \big(2 f_\Lambda^{123} -
f_\Lambda^{134} - f_\Lambda^{234} \big) \Big]\nn[3pt]
&+&
\frac{g^4}{128 F^6} \fet \tau_1 \cdot \fet \tau_2 \fet
\tau_3 \cdot \fet \tau_4 \frac{\vec \sigma_1 \cdot \vec q_1\, \vec \sigma_2 \cdot \vec q_2\, 
  \vec \sigma_3 \cdot \vec q_3\, 
  \vec \sigma_4 \cdot \vec q_4
}{(\vec q_1^{\, 2} + M^2) (\vec q_2^{\, 2} + M^2)  (\vec q_3^{\, 2} + M^2) (\vec q_4^{\, 2} +
  M^2)} \nn[3pt]
&\times & \bigg[- M^2 f_\Lambda + (2 M^2 + \vec q_{12}^{\, 2} )
f_\Lambda^{123} 
 + 2 M^2 (M^2 + \vec q_{1}^{\, 2} ) \frac{f_\Lambda^{134}-f_\Lambda^{234}}{\vec q_{1}^{\, 2} -\vec q_{2}^{\, 2} }
+ 2 (M^2 + \vec q_{2}^{\, 2})  (\vec q_{13}^{\, 2} - \vec q_{12}^{\,
  2} ) \frac{f_\Lambda^{124}-f_\Lambda^{134}}{\vec q_{2}^{\, 2} -\vec
  q_{3}^{\, 2} } \bigg] \nn[3pt]
& + & 23 \text{ perm.},
\label{4NF_HDR}
\eeqa
where we have defined
the regulator functions
\beq
f_\Lambda \,=\, e^{-\frac{\vec q_{1}^{\, 2} + M^2}{\Lambda^2}}
e^{-\frac{\vec q_{2}^{\, 2} + M^2}{\Lambda^2}}
e^{-\frac{\vec q_{3}^{\, 2} + M^2}{\Lambda^2}}
e^{-\frac{\vec q_{4}^{\, 2} + M^2}{\Lambda^2}} , \quad \quad
f_\Lambda^{ijk} \,=\, e^{-\frac{\vec q_{i}^{\, 2} + M^2}{\Lambda^2}}
e^{-\frac{\vec q_{j}^{\, 2} + M^2}{\Lambda^2}}
e^{-\frac{\vec q_{k}^{\, 2} + M^2}{\Lambda^2}}\,.
\eeq
Notice that since $f_\Lambda^{134}-f_\Lambda^{234} = (\vec q_{2}^{\,
  2} - \vec q_{1}^{\, 2} )/\Lambda^2 + \mathcal{O} (\Lambda^{-4})$
and $f_\Lambda^{124}-f_\Lambda^{134} = (\vec q_{3}^{\,
  2} - \vec q_{2}^{\, 2} )/\Lambda^2 + \mathcal{O} (\Lambda^{-4})$,
the regularized expression of the four-nucleon force is nonsingular
for any values of the momentum transfers. We further emphasize that
the obtained result does not depend on the arbitrary
constant $\alpha$ that parametrizes the SU(2) matrix $U$ in terms of
the pion field, see Eq.~(\ref{U_expressed_by_pi}). This nontrivial
feature confirms that the employed higher derivative regularization respects the
chiral symmetry.\footnote{Another implication of the chiral symmetry
  is that the vanishing pion mass in the chiral limit is
  symmetry-protected. Using non-symmetry-preserving cutoff
  regularizations, the tadpole diagram appearing
  in the pion self energy would generally produce $\Lambda^4$-terms
  that contribute to the pion mass in the chiral limit. We have verified
  that no such terms appear using the higher-derivative-regularized
  Lagrangian in Eq.~(\ref{regularizedLpi2Expanded}).}  
Finally, using $f_\Lambda = 1 + \mathcal{O}
(\Lambda^{-2})$ and $f_\Lambda^{ijk} = 1 + \mathcal{O}
(\Lambda^{-2})$, one readily reproduces the unregularized
expression for the considered 4NF given in
Eq.~(3.44) of Ref.~\cite{Epelbaum:2007us} by
taking the limit $\Lambda \to \infty$ in Eq.~(\ref{4NF_HDR}):
\beqa
V^{4N}_{\infty} &=& -\frac{g^4}{64 F^6} \fet \tau_1 \cdot \fet \tau_2 \fet
\tau_3 \cdot \fet \tau_4 \frac{\vec \sigma_2 \cdot \vec q_2\, 
  \vec \sigma_3 \cdot \vec q_3\, 
  \vec \sigma_4 \cdot \vec q_4
}{(\vec q_2^{\, 2} + M^2) (\vec q_3^{\, 2} + M^2) (\vec q_4^{\, 2} +
  M^2)} \, \vec \sigma_1 \cdot \vec q_{12} \nn[3pt]
&+&
\frac{g^4}{128 F^6} \fet \tau_1 \cdot \fet \tau_2 \fet
\tau_3 \cdot \fet \tau_4 \frac{\vec \sigma_1 \cdot \vec q_1\, \vec \sigma_2 \cdot \vec q_2\, 
  \vec \sigma_3 \cdot \vec q_3\, 
  \vec \sigma_4 \cdot \vec q_4
}{(\vec q_1^{\, 2} + M^2) (\vec q_2^{\, 2} + M^2)  (\vec q_3^{\, 2} + M^2) (\vec q_4^{\, 2} +
  M^2)} \big( M^2 + \vec q_{12}^{\, 2} \big)
\,  + \,  23 \text{ perm.},
\label{4NF_NonReg}
\eeqa

It is important to keep in mind that even with the ansatz of
Eq.~(\ref{Lpi2Reg}), there is an infinite number of 
diagrams in the pionic
sector which remain unregularized such as, e.g., the pion tadpole
diagrams. Therefore, one would still have to apply an additional
symmetry-preserving regularization like, e.g., $\zeta$-function
or dimensional regularization on top of the higher derivative one, while
keeping the cutoff $\Lambda$ finite. This would remove the remaining
divergences without introducing any additional ambiguity in
the expressions for the nuclear forces.

\begin{figure}[tb]
  \begin{center} 
\includegraphics[width=0.125\textwidth,keepaspectratio,angle=0,clip]{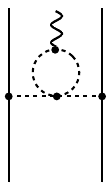}
\caption{Example of a diagram that leads to an exponentially diverging
  loop integral when using higher derivative
  regularization with the ansatz
  of Eq.~(\ref{Lpi2Reg}). Solid, dashed and wavy lines refer to nucleons, 
pions and photons, respectively. Solid dots denote the lowest-order vertices from
the effective chiral Lagrangian. 
  \label{fig:2} 
 }
  \end{center}
\end{figure}

While the considered ansatz of Eq.~(\ref{Lpi2Reg}) solves the
deregularization issue of the naive approach considered in
sec.~\ref{sec:HDRNaive} when applied to nuclear forces, it reaches its
limits when applied to processes involving external sources. For example, the
regularized Lagrangian involving a minimal coupling to a single
photon field $A_\mu^{\rm E}$ 
has the form
\beqa
{\cal L}_{\pi \gamma,\Lambda}^{\rm E \, (2)}&=&-\frac{e}{2}\epsilon_{3ij}
\, A_\mu^{{\rm
    E}}\bigg(\pi^{j} \,
e^{\frac{-\partial^2+M^2}{\Lambda^2}}\partial_\mu\pi^{i}-\big(\partial_\mu\pi^{j}\big)
e^{\frac{-\partial^2+M^2}{\Lambda^2}}
\pi^{i} \, + \, 
\frac{1}{\Lambda^2}\int_0^1
ds\bigg\{\bigg[
e^{s\frac{-\partial^2+M^2}{\Lambda^2}}
(-\partial^2+M^2)\partial_\mu\pi^{i}\bigg]
e^{(1-s)\frac{-\partial^2+M^2}{\Lambda^2}}
\pi^{j}\nn
&-&\bigg[
e^{s\frac{-\partial^2+M^2}{\Lambda^2}}(-\partial^2+M^2)\pi^{i}\bigg]
e^{(1-s)\frac{-\partial^2+M^2}{\Lambda^2}}
\partial_\mu\pi^{j}\bigg\}
\bigg)
\, + \, {\cal{O}} (\pi^4)\,,
\label{higherder_pion_gamma_interaction}
\eeqa
where $e$ denotes the electric charge. 
Clearly, deregularizing factors also appear in the $\gamma \pi
\pi$-vertices.
Consider now the contribution to the electromagnetic current operator
from the diagram shown in 
Fig.~\ref{fig:2}. As we know, each pion propagator acquires an
exponential regulator. On the other hand, each of the four-pion and the $\gamma \pi
\pi$ vertices involve an exponentially growing factor. Let $l_1$ and
$l_2$ denote the four-momenta of pions inside the loop. Because of
momentum conservation, we have $l_2 = l_1 + k$, with $k$ denoting the
photon momentum.
Suppose that both deregularizing factors from the two
vertices act on the momentum $l_1$. Then, the overall effect of the
regulator for the loop contribution is
\beqa
\exp\bigg(\frac{l_1^2+M^2}{\Lambda^2}\bigg)
\exp\bigg(-\frac{l_2^2+M^2}{\Lambda^2}\bigg)&=&\exp\bigg(-\frac{2
  k\cdot l_1 +k^2}{\Lambda^2}\bigg).
\eeqa
With such an exponential factor, the loop contribution cannot be
regularized even using dimensional or $\zeta$-regularization on top of
higher-derivative, except for the vanishing photon momentum $k=0$.
Thus, the ansatz in Eq.~(\ref{Lpi2Reg}) still appears to be too restricted for
our purposes.

\section{Gradient flow regularization}
\def\theequation{\arabic{section}.\arabic{equation}}
\label{sec:GradientFlow}

In the last two sections, the higher derivative method was
applied to regularize the pion propagator. For both considered
modifications of the effective Lagrangian, the failure to
simultaneously regularize 
nuclear forces and currents was caused by
deregularization of vertices induced by the additional
cutoff-dependent terms. An
alternative variant of the higher derivative regularization, which
is free from the deregularization issue, is provided by the
gradient flow method. During the last decade, the Yang-Mills gradient
flow has developed into a powerful tool to address non-perturbative
renormalization aspects in the context of lattice QCD,
see Refs.~\cite{Luscher:2011bx,Luscher:2013cpa,Luscher:2013vga} for
pioneering work along this line. The usefulness of this method for
lattice QCD applications relies on a smoothing effect for the gauge fields caused
by the flow time evolution, which results in an improved or simplified
short-distance behavior of correlation functions evaluated at a
positive flow time.
The idea to use the gradient flow method as a symmetry-preserving
regulator in the context of nuclear chiral EFT has
been raised
in several talks by Kaplan 
\cite{Kaplan}\footnote{To the best of our knowledge, these ideas have not
  been worked out and published elsewhere.}. In the considerations 
below, we follow Ref.~\cite{Kaplan} and show that
the resulting regularization scheme fulfills all our requirements
and allows one to construct consistently regularized nuclear forces and currents.

Consider the lowest-order Lagrangian for pions given in Eq.~(\ref{LagrLO}). 
The corresponding equation of motion is worked out in Appendix
\ref{Derivation_of_EOM} and given in
Eq.~(\ref{EOMDefinition}). To regularize the pion field, we extend the
SU(2) matrix $U(x)$ to $W(x, \tau )$, whose dependence on the parameter
$\tau$ is dictated by the chirally covariant version of the gradient flow equation
\cite{Kaplan}
\beqa
\partial_\tau W&=&-i w\,{\rm EOM}(\tau)\,w\label{gradientFlowDefinition}\,,
\eeqa
subject to the boundary condition $W(x, 0)=U(x)$. Here, $w(x, \tau )$
is defined through the relationship $W = w^2$ and coincides with the matrix
$u$ on the boundary, $w(x,0)=u(x)$, while
\beqa
{\rm EOM}(\tau)&=&\big[D_\mu,
w_\mu\big]+\frac{i}{2}\chi_{-}(\tau)-\frac{i}{4}{\rm
  Tr}\,\chi_{-}(\tau)\label{EOMDefinition_tau} \,,
\eeqa
with 
\beqa
\chi_{\pm}(\tau)&=& w^\dagger\chi w^\dagger \pm w \chi^\dagger w.\label{chiOftauDefinition}
\eeqa
As one can see from Eqs~(\ref{EOMDefinition_tau}) and
(\ref{chiOftauDefinition}), the EOM$(\tau)$ term is constructed out of
the EOM defined in Eq.~(\ref{EOMDefinition}) by
replacing every $u$ field with the $w$ field. The pion-field
transformation properties with respect to chiral
SU$(2)_L\times$SU$(2)_R$ rotations are given by 
\beqa
U (x) &\to& R\,U(x)\,L^\dagger\label{chiral_transf_of_U}\,,
\label{UChiralRotations}
\eeqa
where $L \in$ SU$(2)_L$ and $R \in$ SU$(2)_R$. The covariant form of
the gradient flow equation (\ref{gradientFlowDefinition}), along with
the required boundary condition, guarantees that the field $W (x, \tau
)$ has the same transformation properties at any flow ``time'' $\tau$,  
\beqa
W (x, \tau ) &\to& R\,W (x, \tau )\,L^\dagger \,,
\eeqa
and ensure the unitarity of $W (x, \tau )$. These nontrivial features
are proven  in Appendix~\ref{formal_proofs_gradient_flow}.

As will be shown below, the solution of the gradient flow
equation in terms of the pion fields $\fet \pi (x)$
introduces a Gaussian smearing in
Euclidean space-time, which extends over a
distance $\sim \sqrt{2 \tau}$. To show this we solve the gradient flow
equation (\ref{gradientFlowDefinition}) by performing a
$1/F$ expansion of the generalized pion field.
The most general parametrization of the unitary matrix $W$ can be
written as
\beqa
W&=&1+i{\boldsymbol\tau}\cdot{\boldsymbol\phi}\big(1-\alpha\,\fet
\phi^2\big)-\frac{\fet
  \phi^2}{2}\bigg[1+\bigg(\frac{1}{4}-2\alpha\bigg) \fet
\phi^2\bigg]+{\cal O}(\fet \phi^5)\,,
\eeqa
where $\alpha$ is an arbitrary real constant and the explicit form of
the real dimensionless field $\fet \phi (x, \tau )$
is fixed by the solution of 
Eq.~(\ref{gradientFlowDefinition}). We introduce a metric~\cite{Meetz:1969as}
\beqa
g_{ab}&=&-\frac{1}{2}{\rm Tr}\bigg( W^\dagger \frac{\partial W}{\partial
  \phi_a} W^\dagger \frac{\partial W}{\partial \phi_b}\bigg)\,,
\eeqa 
and multiply both sides of the gradient flow
Eq.~(\ref{gradientFlowDefinition}) by $W^\dagger\partial W/\partial
\phi_a W^\dagger$ from the left. Taking trace we obtain
\beqa
g_{ab} \, \partial_\tau\phi_b&=&\frac{i}{2}{\rm Tr}\bigg[\frac{\partial W}{\partial
\phi_a} w^\dagger {\rm EOM}(\tau) w^\dagger\bigg]\,,\label{gradienFlow_with_Metric}
\eeqa
where we used
\beqa
W^\dagger \partial_\tau W &=&W^\dagger\frac{\partial W}{\partial \phi_a}\partial_\tau\phi_a\,.
\eeqa
Next, we invert Eq.~(\ref{gradienFlow_with_Metric}) to rewrite the gradient
flow equation in a form that is more 
convenient for practical calculations:
\beqa
\partial_\tau \phi_a&=&\frac{i}{2}[g^{-1}]_{ab}\, {\rm Tr}\bigg[\frac{\partial W}{\partial
\phi_b} w^\dagger {\rm EOM}(\tau) w^\dagger\bigg]\,.\label{gradienFlow_Practical}
\eeqa
Using the ansatz for $\fet \phi$ in the form of a power series in $1/F$,
\beqa
\phi_b=\sum_{n=0}^\infty\frac{\phi_b^{(n)}}{F^n}\,,
\eeqa
we obtain a series of recursive differential equations. Restricting
ourselves to terms with at most a single insertion of the external
sources, the lowest-order equation takes the form
\beqa
\big[\partial_\tau-(\partial_\mu^x\partial_\mu^x-M^2)
\big]\phi^{(0)}_b (x,\tau)&=&2
B p_b(x)-\partial_\mu a_{\mu b}(x)\,. \label{gradient_flow_LO}
\eeqa
This
linear inhomogeneous differential
equation can be solved by introducing the retarded Green's function, which fulfills
\beqa
\big[\partial_\tau-(\partial_\mu^x\partial_\mu^x-M^2) \big]G(x-y,\tau-s)&=&\delta(\tau-s)\delta^4(x-y)\,,
\eeqa
and is given by
\beqa
G(x,\tau)&=&\theta(\tau)\int \frac{d^4 q}{(2\pi)^4} e^{-\tau(q^2+M^2)} e^{-i\,q\cdot
  x}\,=\,\frac{\theta(\tau)}{16\pi^2\tau^2} \, e^{-\frac{x^2+4 M^2\tau^2}{4\tau}}\,.
\label{GreensFunction}
\eeqa
Here, $q\cdot x$ denotes an Euclidean product $q\cdot x=q_0
x_0+\vec{q}\cdot\vec{x}$. The solution for $\phi^{(0)}_b (x,\tau)$ with the boundary condition
$\fet \phi^{(0)}(x, 0)=0$ can be written in terms of the Green's
function via
\beqa
\phi^{(0)}(x,\tau)_b&=&\int_0^\tau ds \int d^4 y\, G(x-y,\tau-s)\big(2
B p_b(y)-\partial_\mu a_{\mu b}(y)\big)\,.
\eeqa
Obviously, the above solution fulfills the desired boundary
condition. 
It is instructive to Fourier transform $\fet \phi^{(0)} (x, \tau)$ to momentum space:
\beqa
\tilde{\phi}^{(0)}_b (q,\tau)&=&\int d^4 x\, e^{i q\cdot x}\phi^{(0)}_b(x,\tau)\,=\,\int_0^\tau ds \,e^{-\tau(q^2+M^2)}\big[2
B\tilde{p}_b(q)+i q_\mu \tilde{a}_{\mu b}(q)\big]\,,\label{phi0_momentum_space}
\eeqa
where
\beqa
\tilde{p}_b(q)&=&\int d^4 x\, e^{i q\cdot x} p_b(x),\quad \quad 
\tilde{a}_{\mu b}(q)\,=\,\int d^4 x\, e^{i q\cdot x} a_{\mu b}(x)\,.
\eeqa
Eq.~(\ref{phi0_momentum_space}) can be integrated analytically,
leading to 
\beqa
\tilde{\phi}^{(0)}_b (q,\tau)&=&\frac{1-e^{-\tau(q^2+M^2)}}{q^2+M^2}\big[2
B\tilde{p}_b(q)+i q_\mu \tilde{a}_{\mu b}(q)\big]\,.\label{phi0_momentum_space_final}
\eeqa
From Eq.~(\ref{phi0_momentum_space_final}) we see that
$\tilde{\phi}^{(0)}(q,\tau)$ approaches zero linearly with vanishing
$\tau$. It is also easy
to verify that $\tilde{\phi}^{(0)}(q,\tau)$ satisfies the momentum-space
version of the lowest-order gradient
flow equation
\beqa
\big(\partial_\tau+q^2+M^2\big) \tilde{\phi}^{(0)}_b(q,\tau)&=&B\tilde{p}_b(q)+i q_\mu \tilde{a}_{\mu b}(q)\,,\label{gradient_flow_lo_momentum_space}
\eeqa
using the trivial identity
\beqa
\big(\partial_\tau+q^2+M^2\big) \frac{1-e^{-\tau(q^2+M^2)}}{q^2+M^2}&=&1\,.
\eeqa

To keep the discussion of higher-order contributions to
$\fet {\phi}(x,\tau)$ transparent, we switch off all
external sources from now on (except for the scalar source, whose
value is
set to $m_q$). Then, the gradient flow equation at order $1/F$
takes a linear homogeneous form
\beqa
\big[\partial_\tau-(\partial_\mu^x\partial_\mu^x-M^2) \big]\phi^{(1)}_b(x,\tau)&=&0\,,
\eeqa
with the boundary condition $\fet \phi^{(1)}(x,0)= \fet \pi(x)$, where
the pion field $\fet \pi(x)$ is chosen according to Eq.~(\ref{U_expressed_by_pi}). 
The solution is given by
\beqa
\phi^{(1)}_b(x,\tau)&=&\int d^4y \,G(x-y,\tau) \pi_b(y)\,,
\label{phi_1}
\eeqa
which leads to the momentum-space expression 
\beqa
\tilde{\phi}^{(1)}_b (q,\tau)&=&e^{-\tau(q^2+M^2)}\tilde{\pi}_b(q)\,.\label{gradient_flow_solution_nlo_momentum_space}
\eeqa
This shows that
the gradient flow introduces a Gaussian-type regulator of the pion
field.

At order $1/F^2$, the gradient flow equation has again a homogeneous
form
\beqa
\big[\partial_\tau-(\partial_\mu^x\partial_\mu^x-M^2) \big]\phi^{(2)}_b(x,\tau)&=&0\,,
\eeqa
with the boundary condition $\phi^{(2)}_b(x,0)=0$. The only
solution that satisfies this condition is a trivial one
$\phi^{(2)}_b (x,\tau)=0$, so that there are no contributions to
$\phi_b$ at order $1/F^2$ (in the absence of external sources).

At order $1/F^3$, we obtain the inhomogeneous linear differential equation for
$\fet \phi^{(3)} (x, \tau)$
\beqa
\big[\partial_\tau-(\partial_\mu^x\partial_\mu^x-M^2)
\big]\phi^{(3)}_b (x,\tau)&=&(1-2\alpha)\partial_\mu\fet{\phi}^{(1)}\cdot\partial_\mu\fet{\phi}^{(1)}\phi^{(1)}_b-4\alpha\,\partial_\mu\boldsymbol{\phi}^{(1)}\cdot\boldsymbol{\phi}^{(1)}\partial_\mu\phi^{(1)}_b\nn
&+&\frac{M^2}{2}(1-4\alpha)\boldsymbol{\phi}^{(1)}\cdot \boldsymbol{\phi}^{(1)}\phi^{(1)}_b\,.\label{gradient_flow_eq_n3lo}
\eeqa
The boundary condition
for $\fet \phi^{(3)} (x, 0)$ can be derived by examining the matrix $W (x, \tau)$:
\beqa
W&=&1+\frac{i}{F}\boldsymbol{\tau}\cdot\boldsymbol{\phi^{(1)}}-\frac{1}{2
  F^2}\boldsymbol{\phi}^{(1)}\cdot\boldsymbol{\phi}^{(1)} +
\frac{i}{F^3}\big(\boldsymbol{\tau}\cdot\boldsymbol{\phi}^{(3)}-\alpha\,\boldsymbol{\tau}\cdot\boldsymbol{\phi}^{(1)}\boldsymbol{\phi}^{(1)}\cdot\boldsymbol{\phi}^{(1)}\big)
+ \mathcal{O} \bigg( \frac{1}{F^4} \bigg)\,,
\eeqa
where  we have used that $\fet \phi^{(2)} (x, \tau )=0$. Given that $\phi_b^{(1)}(x,
0)=\pi_b(x)$, we obtain
\beqa
W\Big|_{\tau=0}&=&1+\frac{i}{F}\boldsymbol{\tau}\cdot\boldsymbol{\pi}-\frac{1}{2
  F^2}\boldsymbol{\pi}\cdot\boldsymbol{\pi} +
\frac{i}{F^3}\Big(\boldsymbol{\tau}\cdot\boldsymbol{\phi}^{(3)}\Big|_{\tau=0}-\alpha\,\boldsymbol{\tau}\cdot\boldsymbol{\pi}\boldsymbol{\pi}\cdot\boldsymbol{\pi}\Big)\,.
\label{Temp01}
\eeqa
Using $W|_{\tau=0}=U$ and matching Eq.~(\ref{Temp01}) to
Eq.~(\ref{U_expressed_by_pi}), we finally obtain the boundary
condition 
$\fet \phi^{(3)} (x, 0)=0$. We then
write the solution of Eq.~(\ref{gradient_flow_eq_n3lo}) in the form 
\beqa
\phi^{(3)}_b (x,\tau)&=&\int_0^\tau ds \int d^4 y\,
G(x-y,\tau-s)\bigg[(1-2\alpha)\partial_\mu\fet{\phi}^{(1)}(y,s)\cdot\partial_\mu\fet{\phi}^{(1)}(y,s)\,
\phi^{(1)}_b(y,s)\nn
&-&4\alpha\,\partial_\mu\boldsymbol{\phi}^{(1)}(y,s)\cdot\boldsymbol{\phi}^{(1)}(y,s)\,
\partial_\mu\phi^{(1)}_b(y,s)
+\frac{M^2}{2}(1-4\alpha)\boldsymbol{\phi}^{(1)}(y,s)\cdot
\boldsymbol{\phi}^{(1)}(y,s)\, \phi^{(1)}_b(y,s)\bigg]\,.
\label{phi_3}
\eeqa
The corresponding momentum-space expression is given by 
\beqa
\tilde{\phi}^{(3)}_b(q,\tau)&=&\int\frac{d^4 q_1}{(2\pi)^4}\frac{d^4
  q_2}{(2\pi)^4}\frac{d^4
  q_3}{(2\pi)^4}(2\pi)^4\delta^4 (q-q_1-q_2-q_3)\int_0^\tau ds
\,e^{-(\tau-s)(q^2+M^2)}e^{-s\sum_{j=1}^3(q_j^2+M^2)}\nn
&\times&\bigg[4\alpha \,q_1\cdot q_3-(1-2\alpha)q_1\cdot
q_2+\frac{M^2}{2}(1-4\alpha)\bigg]\tilde{\fet
  \pi}(q_1)\cdot\tilde{\fet \pi}(q_2)\, \tilde{\pi}_b (q_3)\,.
\eeqa
By looking at the regulator
\beqa
\int_0^\tau ds
\,e^{-(\tau-s)(q^2+M^2)}e^{-s\sum_{j=1}^3(q_j^2+M^2)}&=&\frac{e^{-\tau(q^2+M^2)}-e^{-\tau\sum_{j=1}^3(q_j^2+M^2)}}{q_1^2+q_2^2+q_3^2-q^2+2M^2}\,,\label{gradient_flow_regulator_for_three_pions}
\eeqa
one observes that not every pion field gets regularized since the  
first Gaussian regulator in the right-hand side of
Eq.~(\ref{gradient_flow_regulator_for_three_pions}) only acts on the
total pion momentum $q = q_1 + q_2 + q_3$. However, as will
be argued below, this regularization is sufficient for our purposes.  Notice further
that the above expression is non-singular for all values of the
momenta $q_i$ and $q$.

\begin{figure}[tb]
  \begin{center} 
\includegraphics[width=\textwidth,keepaspectratio,angle=0,clip]{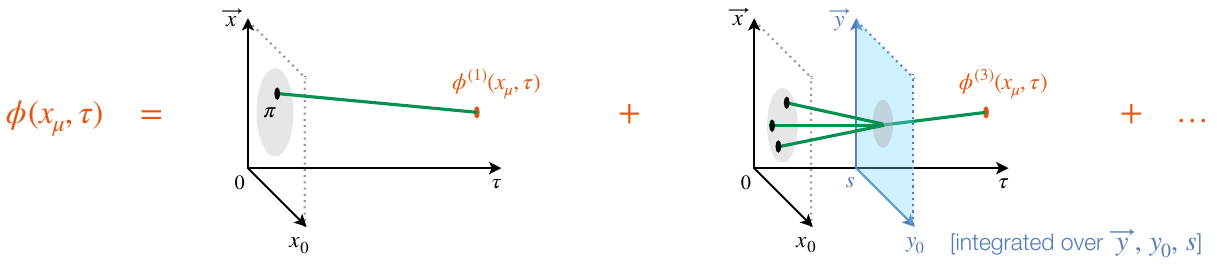}
    \caption{
Schematic graphical representation of the solution of the gradient
flow equation in coordinate space in the absence of external
sources. Red solid dots denote a point in the five-dimensional
$(x_\mu, \tau)$ space, at which the
field $\fet \phi$ is evaluated, while black solid dots on the $\tau=0$
boundary correspond to pion locations (being integrated over). Green solid lines refer
to the Green's function defined in Eq.~(\ref{GreensFunction}). Light
shaded areas visualize schematically the
smearing in Euclidean space-time, whose characteristic size is
governed by the gradient flow time. 
\label{fig:3} 
 }
  \end{center}
\end{figure}

The above considerations help to elucidate the general structure of
the solution of the gradient flow equation  $\fet \phi (x, \tau)$,
which is schematically depicted in Fig.~\ref{fig:3}. Specifically, the
field $\fet \phi (x, \tau)$ is expressed in terms of an increasing
number of smeared pion fields that live on the boundary $\tau = 0$,
with the extent of smearing being controlled by the parameter $\sqrt{2 \tau}$. 
In the limit $\tau \to 0$, all multi-pion contributions to $\fet \phi$
get suppressed and the field $\fet \phi$ turns to the pion field $\fet
\pi$. 

After these preparations, we are now in the position to define our
regularization scheme using the gradient flow method. In the
Goldstone boson sector, we employ the standard (i.e., unregularized)
Lagrangian $\mathcal{L}_\pi = \mathcal{L}_\pi^{(2)} +
\mathcal{L}_\pi^{(4)} + \ldots$, whose lowest-order expression
given in Eq.~(\ref{LagrLO}). The construction of the pionic Lagrangian
is conveniently carried out in terms of the matrix $U$, which
transforms with respect to chiral rotations according to
Eq.~(\ref{UChiralRotations}). Different
parametrizations of $U$ by the pion fields correspond to different
nonlinear realizations of the SU$(2)_L\times$SU$(2)_R$ group, but lead
to the same S-matrix elements. When including matter fields such as
nucleons, it is more convenient to employ the SU$(2)$ matrix $u =
\sqrt{U}$, see also Eq.~(\ref{ChiralBuildingBlocks}), which possesses
somewhat more
complicated transformation properties with 
respect to chiral rotations:
\beqa
u&\to& R\,u\,K^\dagger \,=\,K\,u\, L^\dagger\,,
\eeqa
where the space-time dependent SU$(2)$ matrix $K(L, R, U) =\sqrt{L
  \,U^\dagger R^\dagger} R\, \sqrt{U}$ is referred to as the   
compensator field. Nucleon fields can then be chosen to transform under  
chiral rotations according to $N \, \to \, K N$ \cite{Coleman:1969sm,Callan:1969sn}. The resulting
nonlinear realization of the chiral group allows one to
straightforwardly 
write down the effective chiral Lagrangian in terms of the covariantly
transforming building blocks like $u_\mu \, \to \,  K u_\mu
K^\dagger$, $D_\mu N \, \to \, K D_\mu N$, etc., see
Eq.~(\ref{ChiralBuildingBlocks}) for the relevant definitions. 
To construct the regularized pion-nucleon Lagrangian, we follow the
same path but use  the corresponding building blocks at a
nonzero flow time $\tau$,
i.e., we replace the pion matrix $U$ by $W(\tau )$ in all expressions
in Eq.~(\ref{ChiralBuildingBlocks}) and require the nucleon fields to
transform according to $N \, \to \, K(\tau ) N$, where $K(\tau)=\sqrt{L \,W^\dagger R^\dagger} R\, w$
with $w = \sqrt{W}$\footnote{Notice that for isospin transformations with
$L=R=V$, one has $K(\tau ) = V$, so that our definition is consistent
with nucleons forming an isospin doublet.}. The key feature that enables this construction
and ensures that the resulting Lagrangian is chirally invariant is
the transformation property of the solution of the gradient flow
equation, $W (\tau ) \to R \, W(\tau ) \, L^\dagger$, which as shown  in
Appendix~\ref{formal_proofs_gradient_flow}  holds true for all values of the
flow parameter $\tau$. 

The extension of the Lagrangian to a finite $\tau$, $\mathcal{L}_{\pi N}^{\rm E} \to
\mathcal{L}_{\pi N, \tau}^{\rm E}$, leads to new vertices as visualized in Fig.~\ref{fig:3} and has
a smearing effect on the pion fields that ensures that all pion-nucleon
vertices acquire exponential factors, which decrease with growing
(combinations of) pion momenta. It is also
remarkable that for $\tau \neq 0$, the field $W(\tau)$ does depend, in contrast to the field
$U$, on the external
sources and has a rather complicated structure.
As long as the flow time is kept
sufficiently small, i.e.~$\tau \sim \Lambda_b^{-2} \ll M^{-2}$,
where $\Lambda_b$ is the breakdown scale of the chiral
EFT expansion, the resulting nonlocal effective Lagrangian gives
rise to an EFT description of QCD that is equivalent to
the standard formulation of chiral perturbation theory (albeit with
different values of low-energy constants).

Using the perturbative solution
for $\fet \phi (x, \tau)$ in Eqs.~(\ref{phi_1}) and
(\ref{phi_3}), along with the expression for the Green's function in
Eq.~(\ref{GreensFunction}), the regularized lowest-order pion-nucleon
Lagrangian is found to take the form\footnote{Here, we do not show the
  temporal cutoff in the nucleon kinetic term, which needs to be
  introduced in the derivation of nuclear potentials using the
  path-integral  approach of Ref.~\cite{Krebs:2023ljo} and is removed at the end of the calculation.}
\beqa
\mathcal{L}_{\pi N, \tau}^{\rm E \, (1)} &=& N^\dagger\big(D_0^w+ g
w_\mu S_\mu\big) N\,,
\eeqa
with 
\beqa
D_\mu^w&=&\partial_\mu + \Gamma_\mu^w, \quad \Gamma_\mu^w\,=\, \frac{1}{2}\big[w^\dagger,\partial_\mu
w\big]-\frac{i}{2} w^\dagger r_\mu w - \frac{i}{2} w\, l_\mu
w^\dagger,\quad S_\mu = ( 0, \vec{\sigma}/2)\,.
\eeqa
In the absence of external sources, the regularized Lagrangian $\mathcal{L}_{\pi N,
  \tau}^{\rm E \, (1)}  $, expanded
in powers of the pion field, takes the form
\beqa
{\cal L}_{\pi N, \tau}^{\rm
  E\,(1)}&=&N^\dagger(x)\bigg[-\frac{g}{2F}\vec\sigma\cdot\vec{\nabla}_x{\cal
  G}[\delta\fet\pi](x,\tau)\cdot\fet\tau + \frac{i}{4 F^2}\big(\fet\tau\times{\cal
  G}[\delta\fet\pi](x,\tau)\big)\cdot\partial_0^x{\cal G}[\delta\fet\pi](x,\tau)\nn
&+&\frac{g}{4
  F^3}\vec{\sigma}\cdot\vec{\nabla}_x\bigg(M^2(4\alpha-1){\cal
  G}[\theta{\cal G}[\delta\fet\pi]\cdot{\cal G}[\delta\fet\pi]{\cal
  G}[\delta\fet\pi]](x,\tau)+2(2\alpha-1){\cal G}[\theta {\cal
  G}[\delta\partial_\mu\fet\pi]\cdot {\cal
  G}[\delta\partial_\mu\fet\pi]{\cal
  G}[\delta\fet\pi]](x,\tau)\nn
&+&8\,\alpha\,{\cal G}[\theta{\cal G}[\delta\fet\pi]\cdot{\cal
  G}[\delta\partial_\mu\fet\pi]\cdot{\cal G}[\delta\fet\pi]{\cal
  G}[\delta\partial_\mu\fet\pi]](x,\tau)+2\,\alpha\,{\cal
  G}[\delta\fet\pi](x,\tau)\cdot{\cal G}[\delta\fet\pi](x,\tau) {\cal
  G}[\delta\fet\pi](x,\tau)\bigg)\cdot\fet\tau\nn
&-&\frac{g}{8 F^3}\fet\tau\cdot{\cal
  G}[\delta\fet\pi](x,\tau)\vec{\sigma}\cdot\vec{\nabla}_x\big({\cal
  G}[\delta\fet\pi](x,\tau)\cdot{\cal
  G}[\delta\fet\pi](x,\tau)\big)\bigg]N(x) + {\cal O}(\pi^4)\,.
\label{LagrGFexpanded}
\eeqa
Here, we have introduced the short-hand notation
\beqa
{\cal G}[f](x,\tau)&=&\int_{-\infty}^\tau ds \, e^{-(\tau-s)(-\partial_x^2+M^2)}f(x,s),
\eeqa
where $f(x,s)$ is a continuous function or distribution. In
particular
\beqa
{\cal G}[\delta \fet{\pi}](x,\tau)&=&\int_{-\infty}^\tau  ds \, e^{-(\tau-s)(-\partial_x^2+M^2)}\delta(s)\fet\pi(x)\,=\, e^{-\tau(-\partial_x^2+M^2)}\fet\pi(x)\,,
\eeqa
and
\beqa
{\cal G}[\theta f](x,\tau)&=&\int_{-\infty}^\tau ds \, 
e^{-(\tau-s)(-\partial_x^2+M^2)}\theta(s)f(x,s)\,=\, \int_{0}^\tau ds \, 
e^{-(\tau-s)(-\partial_x^2+M^2)}f(x,s)\,,
\eeqa
such that, e.g.,
\beqa
{\cal G}[\theta {\cal G}[\delta\fet\pi]\cdot {\cal
  G}[\delta\fet\pi]{\cal G}[\delta\fet\pi]](x,\tau)&=&\int_0^\tau ds \, 
e^{-(\tau-s)(-\partial_x^2+M^2)} {\cal G}[\delta\fet\pi](x,s)\cdot
{\cal G}[\delta\fet\pi](x,s)  {\cal G}[\delta\fet\pi](x,s)\\
&=&\int_0^\tau ds \, 
e^{-(\tau-s)(-\partial_x^2+M^2)} \bigg[e^{-s(-\partial_x^2+M^2)}\fet\pi(x)\bigg]\cdot \bigg[e^{-s(-\partial_x^2+M^2)}\fet\pi(x)\bigg]\bigg[e^{-s(-\partial_x^2+M^2)}\fet\pi(x)\bigg]\,.\nonumber
\eeqa
When taking the limit $\tau \to 0$, the regularized expression in
Eq.~(\ref{LagrGFexpanded}) turns into
the unregularized Lagrangian in Eq.~(\ref{LagrPiNNoReg}). 
On the other hand, as already mentioned above, the pionic Lagrangian is not regularized and has
the usual form
\beqa
\mathcal{L}_{\pi}^{\rm E \, (2)} &=& \frac{1}{2} (
\partial_\mu \fet \pi \cdot \partial_\mu \fet \pi  + M^2 \fet \pi^2)
- \frac{1}{F^2} \bigg( \alpha - \frac{1}{8} \bigg) \fet \pi^2 \fet \pi
\cdot (-\partial^2 + M^2) \fet \pi - \frac{1}{4 F^2} \fet \pi^2
\partial_\mu \fet \pi \cdot \partial_\mu \fet \pi - \frac{1 }{8F^2}
\fet \pi^2 \fet \pi \cdot \partial^2 \fet \pi \nn
&+& \mathcal{O} (\fet
\pi^6 )\,.
\eeqa

When using $\mathcal{L}_{\pi N, \tau}^{\rm E \, (1)}$ and
$\mathcal{L}_{\pi}^{\rm E  \, (2)}$ to derive the one-pion exchange
nucleon-nucleon potential, the pion
propagator in the static limit acquires the regulator
\beqa
\frac{1}{\vec q^{\, 2}+M^2}&\to&\frac{e^{-2\tau(\vec q^{\,
      2}+M^2)}}{\vec q^{\, 2}+M^2}\,.
\eeqa
Therefore, to recover the semilocal momentum-space regulator of
Refs.~\cite{Reinert:2017usi,Reinert:2020mcu} we set the flow time to $\tau = 1/ (2 \Lambda^2)$. 

As a proof-of-principle example, we apply the gradient flow method to the
4NF generated by the Feynman diagrams shown in Fig.~\ref{fig:1}. A
straightforward calculation\footnote{It is possible to perturbatively
  evaluate the
  correlation functions at a finite flow time $\tau$ using a local five-dimensional
  quantum field theory that involves additional
  fields associated with Lagrange multipliers needed to ensure the
  proper boundary conditions for the generalized pion fields
  \cite{Luscher:2011bx}.  Here, we follow a different approach
  and perform calculations in four-dimensional Euclidean
  space-time at a fixed $\tau$ using the nonlocal (smeared)
  Lagrangian. The expressions for the diagrams shown in
  Fig.~\ref{fig:1} can be obtained by applying the Feynman technique
  to the nonlocal regularized Lagrangian. More generally, the derivation of
  consistently regularized nuclear forces and currents is going to be carried
  out using the path-integral
  method introduced in Ref.~\cite{Krebs:2023ljo}.} reveals the following result: 
\beqa
V^{4N}_\Lambda &=& \frac{g^4}{64 F^6} \fet \tau_1 \cdot \fet \tau_2 \fet
\tau_3 \cdot \fet \tau_4 \frac{\vec \sigma_2 \cdot \vec q_2\, 
  \vec \sigma_3 \cdot \vec q_3\, 
  \vec \sigma_4 \cdot \vec q_4
}{(\vec q_2^{\, 2} + M^2) (\vec q_3^{\, 2} + M^2) (\vec q_4^{\, 2} +
  M^2)}
\bigg[ \vec \sigma_1 \cdot \vec q_1 \big( 2g_\Lambda - 4
f_\Lambda^{123} + 2 f_\Lambda^{134} - f_\Lambda^{234} \big) - \vec
\sigma_1 \cdot \vec q_2 f_\Lambda^{234}
\nn[3pt]
\nn[3pt]
&& {} \, + 2  \vec \sigma_1 \cdot \vec q_1 \big( 5 M^2 + \vec q_1^{\, 2} + \vec
q_2^{\, 2} + \vec q_3^{\, 2} + \vec q_4^{\, 2} +  \vec q_{34}^{\, 2}
\big)
\frac{g_\Lambda - f_\Lambda^{134}}{2 M^2 + \vec
q_1^{\, 2} + \vec
q_3^{\, 2} + \vec
q_4^{\, 2}  - \vec
q_2^{\, 2} }\nn[3pt]
&& {} \, -   4\vec \sigma_1 \cdot \vec q_1 \big( 3 M^2 + \vec q_1^{\, 2} + \vec
q_2^{\, 2} + \vec q_3^{\, 2} + \vec q_4^{\, 2} -  \vec q_{34}^{\, 2}
\big)
\frac{g_\Lambda - f_\Lambda^{124}}{2 M^2 + \vec
q_1^{\, 2} + \vec
q_2^{\, 2} + \vec
q_4^{\, 2} - \vec
q_3^{\, 2} } \bigg] \\[3pt]
&+&
\frac{g^4}{128 F^6} \fet \tau_1 \cdot \fet \tau_2 \fet
\tau_3 \cdot \fet \tau_4 \frac{\vec \sigma_1 \cdot \vec q_1\, \vec \sigma_2 \cdot \vec q_2\, 
  \vec \sigma_3 \cdot \vec q_3\, 
  \vec \sigma_4 \cdot \vec q_4
}{(\vec q_1^{\, 2} + M^2) (\vec q_2^{\, 2} + M^2)  (\vec q_3^{\, 2} + M^2) (\vec q_4^{\, 2} +
  M^2)} \big( M^2 + \vec q_{12}^{\, 2} \big) \big( 4  f_\Lambda^{123}
- 3 g_\Lambda \big) \, + \, 23 \text{ perm.}, \nonumber
\eeqa
where we have defined 
\beq
g_\Lambda \,=\, e^{-\frac{\vec q_{1}^{\, 2} + M^2}{2\Lambda^2}}
e^{-\frac{\vec q_{2}^{\, 2} + M^2}{2\Lambda^2}}
e^{-\frac{\vec q_{3}^{\, 2} + M^2}{2\Lambda^2}}
e^{-\frac{\vec q_{4}^{\, 2} + M^2}{2\Lambda^2}}\,.
\eeq
The expression obtained using the gradient flow method
differs from the one in Eq.~(\ref{4NF_HDR}) calculated using the higher-derivative
regularization ansatz of Eq.~(\ref{Lpi2Reg}), but it also reduces to
Eq.~(\ref{4NF_NonReg}) in the $\Lambda \to \infty$ limit and shows no dependence
on the arbitrary parameter $\alpha$, which provides a nontrivial
check of our results.   

\begin{figure}[tb]
  \begin{center} 
\includegraphics[width=0.65\textwidth,keepaspectratio,angle=0,clip]{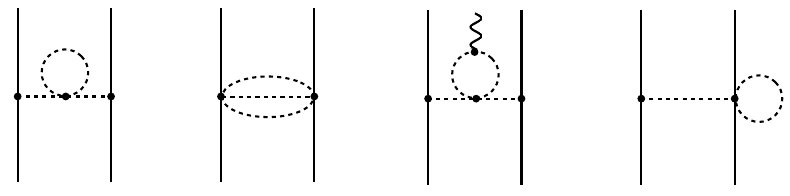}
\caption{Examples of contributions to the nuclear forces and currents
  that involve loop integrals, which are not regularized in the gradient flow
  method. Solid, dashed and wavy lines refer to nucleons, 
pions and photons, respectively. Solid dots denote the lowest-order vertices from
the effective chiral Lagrangian. 
  \label{fig:4} 
 }
  \end{center}
\end{figure}

Similarly to the higher-derivative regularization considered in
sec.~\ref{sec:HDR}, the gradient flow
method does not eliminate all ultraviolet divergences that appear in
loop contributions to the nuclear forces and currents. Some examples of
the divergent loop diagrams are shown in Fig.~\ref{fig:4}. 
However, all such loop contributions involve only pion physics and do not depend
on the nucleon propagators. They can, therefore, be safely regularized using
an additional dimensional or $\zeta$-function
regularization without running into the inconsistency issues mentioned
in the introduction. Crucially, all resulting long-range potentials are guaranteed to be sufficiently regularized
with respect to the external nucleon momenta, i.e., they can be
directly employed
in the many-body Schr\"odinger equation. A distinctive
feature of the gradient-flow regularization method is that it does, per construction, not
lead to any exponential enhancement in the vertices. For this
reason, no issues with exponentially growing integrands that plague
the application of the higher-derivative regularization of sec.~\ref{sec:HDR} to
processes involving external sources appear in the gradient-flow
method. We thus conclude that this approach fulfills all requirements for a
symmetry-preserving regularization scheme formulated in the
Introduction.

\section{Summary and conclusions}
\def\theequation{\arabic{section}.\arabic{equation}}
\label{sec:summary}

In this paper, we have discussed different ways of introducing a cutoff
regulator in chiral EFT for nuclear systems. The regulator is required
to provide a sufficient suppression of nuclear potentials
mediated by single and multiple pion exchanges at large momenta to enable their usage in
the many-body Schr\"odinger equation. In addition, it must
respect the chiral and gauge symmetries, and it should
ideally reduce to a Gaussian-type cutoff of Refs.~\cite{Reinert:2017usi,Reinert:2020mcu} when
applied to the one-pion exchange two-nucleon potential. 

A convenient way to introduce a symmetry-preserving cutoff
regulator in an EFT is to appropriately modify the effective
Lagrangian by including certain higher-order interactions, as
suggested a long time ago by Slavnov in the context of the nonlinear
$\sigma$-model \cite{Slavnov:1971aw}. In our first approach discussed in
sec.~\ref{sec:HDRNaive}, we have considered a simple modification of the
lowest-order effective Lagrangian for pions by replacing ${\rm Tr} \big[
\partial_\mu U^\dagger \partial_\mu U \big]\, \to  \, {\rm Tr} \Big[
\partial_\mu U^\dagger e^{- \partial^2/\Lambda^2}\partial_\mu U
\Big]$.  This modification improves the ultraviolet behavior of the
pion propagator, but at the same time introduces new exponentially
enhanced vertices with four and more pion fields, which have
deregularizing effects. Moreover, since the differential operator in
the exponent of the modified Lagrangian acts on several pion fields, 
the deregularizing factors that appear in the potentials depend
on linear combinations of pion momenta and tend to compensate several
suppressing factors generated by the regularized pion
propagators. In particular, we have shown that the 4NF corresponding to the second diagram
of Fig.~\ref{fig:1} is not sufficiently regularized using the modified
Lagrangian of Eq.~(\ref{Lpi2LambdaTrial}). 

The above-mentioned issue can be mitigated if the additional
cutoff-dependent terms in
the Lagrangian are taken to be proportional to the
equation-of-motion terms. Indeed, using the more sophisticated modification of
the pion Lagrangian specified in Eq.~(\ref{Lpi2Reg}) and performing
an expansion in powers of the pion fields, all deregularizing exponential operators
are found to act on a single pion field, rendering the considered 4NF 
sufficiently regularized. While the modified Lagrangian of
Eq.~(\ref{Lpi2Reg}) can be applied  to derive regularized expressions
for long-range nuclear potentials, it still faces its limitations when
applied to processes involving external sources. For example, the contribution to the
electromagnetic two-nucleon current operator from the diagram shown in
Fig.~\ref{fig:2} involves an ill-defined  loop integral with an exponentially
growing integrand (for nonvanishing values of the photon momentum). 
The considered ansatz is, therefore, still too restricted for
our purposes. 

The last approach we have considered is the gradient flow method \cite{Luscher:2011bx,Luscher:2013cpa,Luscher:2013vga}. Following
Ref.~\cite{Kaplan}, we have introduced a generalized SU(2) matrix $W$ that
depends on the artificial flow time $\tau$ and reduces to the ordinary
pion matrix $U$ on the $\tau = 0$-boundary, $W
\stackrel {\tau \to 0}{\longrightarrow} U$. The $\tau$-evolution of
$W$ is governed by the chirally covariant form of the gradient flow
equation (\ref{gradientFlowDefinition}). We have explicitly shown that this equation
maintains the transformation properties of the matrix $W$ with respect
to chiral rotations, i.e.~the relationship $W \to R W L^\dagger$ holds true for all
values of $\tau$. The matrix $W$ can be expressed in terms of the
ordinary pion field $U$ by solving the gradient flow equation, which
leads to the appearance of new vertices and Gaussian smearing factors that
regularize one or several pion fields. Differently to the previously considered
schemes, regularization is achieved by extending the pion-nucleon
Lagrangian to a finite flow time $\tau$ while keeping the pionic Lagrangian
 in its original, unregularized form. This ensures that no
problematic exponentially increasing factors emerge (also in the presence of
external sources). Moreover, $A$-nucleon connected diagrams acquire
exponentially decreasing factors in at least $A-1$ independent
combinations of the nucleon momentum transfers $\vec q_i$, leading to
sufficiently regularized nuclear forces and currents. As an explicit
example, we worked out the regularized expressions for the 4NF corresponding to the
diagrams shown in Fig.~\ref{fig:1}. The gradient flow method is found to
comply with all requirements we impose on the regularization scheme. 

In the current paper we have limited ourselves to the application of the regularized effective
Lagrangian to the 4NF
shown in Fig.~\ref{fig:1}.
These diagrams are of special interest as they involve non-linear
pion and pion-nucleon vertices constrained by the chiral
symmetry, thereby offering  a non-trivial testing ground of the symmetry-preserving
nature of the considered regularizations. Since these tree-level
diagrams do not involve reducible topologies, the
corresponding 4NF can be read off from the scattering amplitude that
can be obtained using the Feynman rules. For more general types of diagrams
that involve reducible topologies, derivation of nuclear forces and
currents requires separating out the irreducible part of the
amplitude, which goes beyond the Feynman calculus. In Ref.~\cite{Krebs:2023ljo}, we
have formulated a general method to derive nuclear interactions using
the path-integral approach, which is capable of treating regularized
Lagrangians that involve arbitrarily high number of time derivatives. In
subsequent publications, we will apply this method to the
gradient-flow-regularized effective Lagrangian to derive consistent
nuclear forces and current operators.

\section*{Acknowledgments}

We are grateful to Vadim Baru, Arseniy Filin, Ashot Gasparyan, Jambul
Gegelia and Ulf~Mei{\ss}ner for useful comments on the manuscript. 
We also thank all members of the LENPIC collaboration for
sharing their insights into the considered topics.
This work is supported in part
by the European Research Council (ERC) under the EU
Horizon 2020 research and innovation programme (ERC AdG
NuclearTheory, grant agreement No. 885150), by DFG and NSFC through
funds provided to the Sino-German CRC 110 ``Symmetries and the
Emergence of Structure in QCD'' (DFG Project ID 196253076 - TRR 110, NSFC Grant No. 11621131001), by the MKW NRW under the funding code NW21-024-A and by
the EU Horizon 2020 research and innovation programme (STRONG-2020,
grant agreement No. 824093).

\appendix
\renewcommand{\theequation}{\thesection.\arabic{equation}}

\section{Derivation of the 4NF in Eq.~(\ref{4NFNaive}) using the
  path-integral approach of Ref.~\cite{Krebs:2023ljo}}
\label{Derivation_of_4NF}

In Ref.~\cite{Krebs:2023ljo}, we have shown how
to apply the path-integral approach to the derivation of 
nuclear potentials. While the 4NF 
in Eq.~(\ref{4NFNaive}) can be easily obtained by calculating the
corresponding Feynman diagram as done in sec.~\ref{sec:HDRNaive}, here we show how this result
can be recapitulated using the path-integral method of
Ref.~\cite{Krebs:2023ljo}. Following that work, we consider the
lowest-order regularized action for interacting pions
and nucleons,
\beqa
\label{Action1Example}
S^{\rm E}&=&\int d^4 x\, {\cal L}^{\rm E}_{\Lambda},\\
{\cal L}^{\rm E}_\Lambda&=&N^\dagger\bigg[\frac{\partial}{\partial
  x_0}\left(1+\frac{1}{\Lambda_\tau}\frac{\partial}{\partial
    x_0}\right)-\frac{\vec{\nabla}^{\,2}}{2 m}-\frac{g}{2
  F}\vec{\sigma}\cdot\vec{\nabla} \,\fet{\pi}\cdot\fet{\tau}\bigg]N \,
-\, 
  \frac{1}{2}\fet{\pi}\cdot \partial^2
    e^{-\partial^2/\Lambda^2}\fet{\pi}
\, +\, \frac{1}{8
   F^2}\fet \pi^2 \,\partial^2e^{-\partial^2/\Lambda^2}
 \fet \pi^2,
 \nonumber
 \eeqa
where $N$ and $m$ denote the nucleon field and mass, respectively,
$\vec \sigma$ are the Pauli spin matrices and $g$ refers to the bare axial
vector coupling constant in the chiral limit. Further, $\Lambda_\tau$
is a temporal cutoff introduced for technical reasons that will be
removed at the end of the calculation by taking the limit
$\Lambda_\tau \to \infty$, see  Ref.~\cite{Krebs:2023ljo} for details.
The Weinberg-Tomozawa and three-pion-nucleon vertices are not relevant
for the illustrative purposes of this example and have been
dropped. The action in Eq.~(\ref{Action1Example}) coincides with the one
of the Yukawa model considered in Ref.~\cite{Krebs:2023ljo} except
for the vanishing pion mass and the additional four-pion interaction term.  
Following Ref.~\cite{Krebs:2023ljo}, we proceed with  performing
the functional integration over pion fields. Since
Eq.~(\ref{Lpi2ChiralLimitExpandedSigmaGauge}) is not quadratic in
$\fet \pi (x)$, we can integrate over the pion field only approximately by using
the saddle-point method, i.e.~we expand around the minimum of the
action. This is equivalent to the loop expansion,
which is consistent with the chiral expansion.
To find the classical field configuration that minimizes the action $S^{\rm E}$, we take a
functional derivative in the pion field and equate it to zero:
\beqa
\frac{\delta S^{\rm E}}{\delta \pi^j(x)}&=&\frac{g}{2
  F}\vec{\nabla}\cdot\left(N^\dagger(x)\vec{\sigma}[\fet{\tau}]^j
  N(x)\right) \, -\, \partial^2
    e^{-\partial^2/\Lambda^2} \pi^j(x)\, +\, \frac{1}{2
   F^2}\pi^j(x) \, \partial^2 e^{-\partial^2/\Lambda^2}
\big(\fet{\pi} (x) \big)^2 
\, = \, 0.\label{EOMChiralLimit}
\eeqa
The formal solution of this equation of motion  for the pion field is
given by
\beqa
\fet{\pi}^{c}_1&=&\frac{g}{2F} N_2^\dagger \fet{\tau} \sigma^i
N_2 \big(\nabla^i_2\Delta^{\rm
E}_{12}\big)
\, +\, \frac{1}{2
  F^2}\Delta_{12}^{\rm E} \, \fet{\pi}^{c}_2 \, \big(\Delta^{\rm
  E}_{23} \big)^{-1} \, \fet{\pi}^{c}_3\cdot\fet{\pi}^{c}_3 \, + \, {\cal O}(\pi^5)\,,
\eeqa
where $N_i \equiv N (x_i)$, $\fet
\pi_i^c \equiv \fet \pi^c (x_i)$, $\vec \nabla_i \equiv \vec \nabla_{x_i}$ and the integrations over the
coordinates $x_2$ and $x_3$ are assumed. Further, $\Delta_{ij}^{\rm
  E} \equiv \Delta^{\rm E} (x_i - x_j)$ denotes the corresponding Green's function (i.e., the
regularized Euclidean pion propagator in the chiral limit as defined
in Eq.~(\ref{PionPropCL})). 
Replacing the pion field $\fet{\pi}$ in Eq.~(\ref{EOMChiralLimit}) by the classical solution
$\fet{\pi}^c$, expressed in terms of the nucleon fields, we obtain the nucleonic action in the form
\beqa
S^{\rm E}_N&=&N^\dagger_1 \bigg[ \partial_1^0\bigg(1+\frac{\partial_1^0}{\Lambda_\tau}\bigg)
  -\frac{\vec{\nabla}^{\,2}}{2 m}\bigg] N_1 \, +\, \frac{g^2}{8
    F^2} \fet{\tau}_1\cdot\fet{\tau}_2 \big(
  \vec{\sigma}_1\cdot\vec{\nabla}_1 \vec{\sigma}_2\cdot\vec{\nabla}_1
  \Delta^{\rm E}_{12} \big)\label{TrialRegularizationResult}\\
&-&\frac{g^4}{128 F^6}
\fet{\tau}_1\cdot\fet{\tau}_2\fet{\tau}_3\cdot\fet{\tau}_4
\big(
\vec{\sigma}_1\cdot\vec{\nabla}_1 \Delta^{\rm E}_{15} \big)
\big(
\vec{\sigma}_2\cdot\vec{\nabla}_2 \Delta^{\rm E}_{25} \big)
\big(
\vec{\sigma}_3\cdot\vec{\nabla}_3 \Delta^{\rm E}_{36} \big)
\big(
\vec{\sigma}_4\cdot\vec{\nabla}_4 \Delta^{\rm E}_{46} \big)
\big(\Delta^{\rm
  E}_{56} \big)^{-1}\,,\nonumber
\eeqa
where we use the short-hand notation of Ref.~\cite{Krebs:2023ljo}. 
In particular, we suppress
the nucleon fields
$N_i$ with $i=1, \ldots, 4$
(except for the kinetic term) and do not indicate explicitly the integrations over the coordinates
$x_1, \ldots , x_6$. The subscript $i$ of the Pauli spin and isospin
matrices indicates that they are evaluated between the nucleon fields
$N^\dagger (x_i)$ and $N (x_i)$.
This is certainly not the full result since we have neglected
Weinberg-Tomozawa and three-pion-nucleon interactions. For this reason,
the result in Eq.~(\ref{TrialRegularizationResult}) is only valid in the $\sigma$-model parametrization of the
matrix $U$ specified in Eq.~(\ref{SigmaModG}).
While one still needs
to bring the action $S^{\rm E}_N$ into the instantaneous form
as described in Ref.~\cite{Krebs:2023ljo}, the
contribution to the four-nucleon force we are interested in here
can be directly read off from the last term in 
Eq.~(\ref{TrialRegularizationResult}) by taking the functional
derivatives with respect to the nucleon fields.
Performing the Fourier
transformation to momentum space, the corresponding contribution to the 4N
force in the static limit, i.e.~for $q_j^0=0$ with $j=1,\dots,4$, takes
the form given in Eq.~(\ref{4NFNaive}).

\section{Equation of motion for the pion field}
\label{Derivation_of_EOM}

In this appendix we sketch the derivation of the equation of
motion for pion fields, following
the steps described in Ref.~\cite{Scherer:2002tk}, see also the
textbook \cite{Meissner:2022cbi}. We start with the lowest-order
pion Lagrangian in Eq.~(\ref{LagrLO})
and consider the action
\beqa
S^{(2)}&=&\int d^4 x \frac{F^2}{4}{\rm Tr}\big[\big(\nabla_\mu
U\big)^\dagger \nabla_\mu U- U^\dagger\chi-\chi^\dagger U\big]\,.
\eeqa
We perform a variation of the field $U$ by
\beqa
U&\to& e^{i{\fet\epsilon}\cdot{ \fet\tau}}
U\quad\Rightarrow\quad \delta U\,=\,i{\fet \epsilon}\cdot{\fet \tau} U.
\eeqa
For the variation of the action $S^{(2)}$, we obtain
\beqa
\delta S^{(2)}&=&\int d^4 x \frac{F^2}{4}{\rm Tr}\big[-\delta
U^\dagger\nabla_\mu\nabla_\mu U
- (\nabla_\mu\nabla_\mu U)^\dagger \delta U- \delta
U^\dagger\chi-\chi^\dagger \delta U\big]\nn
&=&\int d^4 x \frac{F^2}{4}{\rm Tr}\big[i\,{\fet \epsilon}\cdot{\fet
  \tau}\big((\nabla_\mu\nabla_\mu U)U^\dagger-U(\nabla_\mu\nabla_\mu
U)^\dagger+\chi U^\dagger -  U\chi^\dagger\big)\big]\,.
\eeqa
Accordingly, we obtain for the classical equation of motion
\beqa
{\rm Tr}\big[
  \tau_a\big((\nabla_\mu\nabla_\mu U)U^\dagger-U(\nabla_\mu\nabla_\mu
U)^\dagger+\chi U^\dagger - U \chi^\dagger \big)\big]&=&0\,.\label{EOM_Simpler_Version}
\eeqa
We can collect these three equations into a single matrix
by noting that any $2\times 2$ matrix $A$ can be written in form
\beqa
A&=&\mathbb{1} \frac{1}{2}{\rm Tr} A +\frac{1}{2} {\fet \tau}\cdot{\rm
  Tr}\big({\fet \tau} A\big)\,.
\eeqa
In order for $A=0$ to be equivalent to ${\rm Tr}(\tau_a A)=0$, the matrix
$A$ has to be traceless. Applying this to the equation of motion
(\ref{EOM_Simpler_Version}) we obtain
\beqa
(\nabla_\mu\nabla_\mu U)U^\dagger-U(\nabla_\mu\nabla_\mu
U)^\dagger+\chi U^\dagger - U\chi^\dagger \ - \frac{1}{2}{\rm Tr}\big[(\nabla_\mu\nabla_\mu U)U^\dagger-U(\nabla_\mu\nabla_\mu
U)^\dagger+\chi U^\dagger -U \chi^\dagger \big] &=&0,\label{EOM_First_Step}
\eeqa
where we made the equation of motion operator traceless by subtracting a
half of its trace. Eq.~(\ref{EOM_First_Step}) can be simplified 
using the relationship
\beqa
{\rm Tr}\big(\nabla_\mu\nabla_\mu U U^\dagger\big)&=&{\rm
  Tr}\big[U(\nabla_\mu\nabla_\mu U)^\dagger\big]\,.\label{Trace_Relation}
\eeqa
This equality relies on the fact that both $r_\mu$ and $l_\mu$ are traceless
and that $\partial_\mu U U^\dagger$ is
an element of the Lie algebra, i.e.~it is traceless. Applying the derivative operator
on ${\rm Tr}(\partial_\mu U U^\dagger-U\partial_\mu U^\dagger)=0$ one gets
${\rm Tr}(\partial_\mu\partial_\mu U U^\dagger)={\rm
  Tr}(U \partial_\mu\partial_\mu U^\dagger)$. Rearranging the source terms we
obtain Eq.~(\ref{Trace_Relation}).
Using this relationship Eq.~(\ref{EOM_First_Step}) 
simplifies to 
\beqa
(\nabla_\mu\nabla_\mu U)U^\dagger-U(\nabla_\mu\nabla_\mu
U)^\dagger+\chi U^\dagger - U\chi^\dagger  - \frac{1}{2}{\rm Tr}\big[\chi U^\dagger - U\chi^\dagger \big] &=&0.\label{EOM_Second_Step}
\eeqa
To bring the equation of motion to an even more compact form we multiply both
sides of Eq.~(\ref{EOM_Second_Step}) with $u$ from the right and with
$u^\dagger$ from the left as well as with a prefactor $i/2$:
\beqa
{\rm EOM}\,=\,\big[D_\mu,u_\mu\big]+\frac{i}{2}\chi_- - \frac{i}{4}{\rm Tr}\chi _-&=&0\,.\label{EOM_Final_Step}
\eeqa
Here we used the relationship
\beqa
\frac{i}{2} \big(u^\dagger(\nabla_\mu\nabla_\mu U) u^\dagger-u(\nabla_\mu\nabla_\mu
U)^\dagger u\big)&=&\big[D_\mu, u_\mu\big]\,.
\eeqa

\section{Derivation of the regularized Lagrangian 
in Eq.~(\ref{regularizedLpi2Expanded})}
\label{AppRegularizedLagrangian}

In this appendix we expand the regularized Lagrangian of
Eq.~(\ref{Lpi2Reg}) in pion fields $\fet{\pi}$. We
expand the matrix $u_\mu$ in powers of the pion field
\beqa
u_\mu&=&-\frac{\fet{\tau}\cdot\partial_\mu\fet{\pi}}{F}-\frac{\fet{\tau}\cdot\fet{\pi}\fet{\pi}\cdot\partial_\mu\fet{\pi}}{2
F^3}+\frac{\alpha}{F^3}\Big(\fet{\pi}^2\fet{\tau}\cdot\partial_\mu\fet{\pi}
+2
\fet{\tau}\cdot\fet{\pi}\fet{\pi}\cdot\partial_\mu\fet{\pi}\Big)+{\cal
  O}(\fet \pi^5),\label{alphaChiral}
\eeqa
where $\alpha$ is an arbitrary parameter that appears in the
parametrization of the ${\rm SU}(2)$ matrix
$U$, see Eq.~(\ref{U_expressed_by_pi}).
In Eq.~(\ref{alphaChiral}) we set  all external sources $l_\mu$ and
$r_\mu$ to zero since we focus here on nuclear forces and do not discuss
electroweak processes. From Eq.~(\ref{alphaChiral}), we derive the
expansion of the ${\rm EOM}$ in powers of the pion field:
\beqa
{\rm
  EOM}&=&\big(-\partial^2+M^2\big)\bigg[\frac{\fet{\tau}\cdot\fet{\pi}}{F}\bigg(1-\frac{\alpha}{F^2}\fet{\pi}^2\bigg)\bigg]-\frac{\fet{\tau}\cdot\fet{\pi}}{F^3}\bigg(\partial_\mu\fet{\pi}\cdot\partial_\mu\fet{\pi}+\frac{1}{2}\fet{\pi}\cdot\partial^2\fet{\pi}\bigg)
+ {\cal O}(\fet \pi^5).\quad
\eeqa
To expand  Eq.~(\ref{Lpi2Reg}) in powers of $\fet \pi$, we first
need to obtain an expansion of the exponential operator, which depends
on the covariant derivatives. To do this we introduce a modified covariant
derivative
\beqa
D_\mu^\lambda&=&\partial_\mu + \lambda \Gamma_\mu,
\eeqa
where $\lambda$ is an arbitrary parameter. For $\lambda=0$, the
connection $\Gamma_\mu$ is switched off while for $\lambda=1$, the
covariant derivative acquires its original form. In a similar way, we
introduce
\beqa
\chi_+^\lambda&=&2 M^2 +\lambda \left(\chi_+-2M^2\right).
\eeqa
Furthermore, we use the expressions
\beqa
\Gamma_\mu&=&\frac{i}{4
  F^2}\fet{\tau}\cdot(\fet{\pi}\times\partial_\mu\fet{\pi})+\frac{i}{16
F^4}\fet{\pi}^2\fet{\tau}\cdot(\fet{\pi}\times\partial_\mu\fet{\pi})-\frac{i\,\alpha}{2
F^4}\fet{\pi}^2\fet{\tau}\cdot(\fet{\pi}\times\partial_\mu\fet{\pi})+{\cal
O}(\fet \pi^6),\nn
\chi_+&=&2 M^2-\frac{M^2}{F^2}\fet{\pi}^2-\frac{(1-8\alpha) M^2}{4
  F^4} \fet{\pi}^4+{\cal O}(\fet \pi^6)\,,
\eeqa
where we have switched off all electroweak and pseudoscalar external
sources and set the scalar source to the light quark mass $m_q$. Isospin-breaking
corrections to $\chi_+$
are neglected. The exponential operator in Eq.~(\ref{Lpi2Reg}) can be expressed in the form
\beqa
\exp\left(\frac{-{\rm ad}_{D_\mu}{\rm
      ad}_{D_\mu}+\frac{1}{2}\chi_+}{\Lambda^2}\right)&=&\exp\left(\frac{-{\rm
    ad}_{\partial_\mu}{\rm ad}_{\partial_\mu}+M^2}{\Lambda^2}\right)\bigg[1
+\int_0^1 ds\, \exp\bigg(-s \frac{-{\rm
    ad}_{\partial_\mu}{\rm ad}_{\partial_\mu}+M^2}{\Lambda^2}\bigg)
\nn
&\times&\frac{-\left\{{\rm
    ad}_{\partial_\mu},{\rm ad}_{\Gamma_\mu}\right\}+\frac{1}{2}\big(\chi_+-2M^2\big)}{\Lambda^2}\exp\bigg(s \frac{-{\rm
    ad}_{\partial_\mu}{\rm
    ad}_{\partial_\mu}+M^2}{\Lambda^2}\bigg)\bigg] +{\cal O}(\fet \pi^4).\label{PartialInLambda}
\eeqa
The identity in Eq.~(\ref{PartialInLambda}) is easily proven by using
time-dependent perturbation theory. For the sake of compactness, we
introduce 
\beqa
B&=&\frac{-{\rm ad}_{D_\mu}{\rm
      ad}_{D_\mu}+\frac{1}{2}\chi_+}{\Lambda^2},\quad \quad B_0\,=\,\frac{-{\rm
      ad}_{\partial_\mu}{\rm ad}_{\partial_\mu}+M^2}{\Lambda^2},
  \quad \quad
  \delta B\,=\, B-B_0\,=\, \frac{-\left\{{\rm
    ad}_{\partial_\mu},{\rm ad}_{\Gamma_\mu}\right\}+\frac{1}{2}\big(\chi_+-2M^2\big)}{\Lambda^2}.
 \nonumber
\eeqa
We recall the well-known result of time-dependent perturbation theory 
\beqa
\exp( B)&=&\exp(B_0) T\exp\bigg(\int_0^1 ds \,\delta
B_I(s)\bigg),
\eeqa
where the perturbation in the interaction picture is given by
\beqa
\delta B_I(s)&=&\exp(-s B_0)\delta B\exp(s B_0).
\eeqa
Expanding the operator
\beqa
T\exp\bigg(\int_0^1 ds \,\delta
B_I(s)\bigg)&=&1+\sum_{n=1}^\infty\int_0^1 ds_1\int_0^{s_1}
ds_2\dots\int_0^{s_{n-1}} ds_n \delta B_I(s_1) \delta B_I(s_2)\dots\delta B_I(s_n),
\eeqa
up to first order in $\delta B_I(s)$ yields Eq.~(\ref{PartialInLambda}).
Notice that we do not
employ the large-cutoff expansion since $\Lambda$
is going to be kept fixed throughout the calculation. The only
expansion we make use of is the chiral
expansion in powers of $1/(4\pi F)$. Using Eq.~(\ref{PartialInLambda}), we obtain the
expression for the operator appearing in the Lagrangian in the form
\beqa
\frac{1-\exp\left(\frac{-{\rm ad}_{D_\mu}{\rm
      ad}_{D_\mu}+\frac{1}{2}\chi_+}{\Lambda^2}\right)}{-{\rm ad}_{D_\mu}{\rm
      ad}_{D_\mu}+\frac{1}{2}\chi_+}&=&\frac{1-\exp\Big(\frac{-{\rm
    ad}_{\partial_\mu}{\rm ad}_{\partial_\mu}+M^2}{\Lambda^2}\Big)}{-{\rm
    ad}_{\partial_\mu}{\rm ad}_{\partial_\mu}+M^2} + K_1 + K_2 +{\cal
  O}(\fet \pi^3),
\eeqa
where the operators $K_{1,2}$ are defined by
\beqa
K_1&=&-\frac{1}{-{\rm ad}_{\partial_\mu}{\rm ad}_{\partial_\mu}+M^2}\Big[-\big\{{\rm ad}_{\partial_\mu},{\rm
  ad}_{\Gamma_\mu}\big\}+\frac{1}{2}\big(\chi_+-2 M^2\big)\Big]
\frac{1}{-{\rm ad}_{\partial_\mu}{\rm
    ad}_{\partial_\mu}+M^2}\bigg[1-\exp\bigg(\frac{-{\rm
    ad}_{\partial_\mu}{\rm
    ad}_{\partial_\mu}+M^2}{\Lambda^2}\bigg)\bigg],\nn
K_2&=&-\frac{1}{-{\rm
    ad}_{\partial_\mu}{\rm ad}_{\partial_\mu}+M^2}\int_0^1 ds\exp\bigg[(1-s) \frac{-{\rm ad}_{\partial_\mu}{\rm
      ad}_{\partial_\mu}+M^2}{\Lambda^2}\bigg]\frac{-\big\{{\rm ad}_{\partial_\mu},{\rm
  ad}_{\Gamma_\mu}\big\}+\frac{1}{2}\big(\chi_+-2
M^2\big)}{\Lambda^2}\nn
&\times&\exp\bigg(s \frac{-{\rm ad}_{\partial_\mu}{\rm
      ad}_{\partial_\mu}+M^2}{\Lambda^2}\bigg).
\eeqa
Here, we used the standard relation
\beqa
\frac{1}{A+B}-\frac{1}{A}&=&-\frac{1}{A}B\frac{1}{A}+{\cal O}(B^2),
\eeqa
valid for operators $A$ and $B$. We can rewrite $K_1$ into
\beqa
K_1&=&\frac{1}{-{\rm ad}_{\partial_\mu}{\rm ad}_{\partial_\mu}+M^2}\frac{-\big\{{\rm ad}_{\partial_\mu},{\rm
  ad}_{\Gamma_\mu}\big\}+\frac{1}{2}\big(\chi_+-2
M^2\big)}{\Lambda^2}\int_0^1 ds \exp\bigg(s\frac{-{\rm
    ad}_{\partial_\mu}{\rm
    ad}_{\partial_\mu}+M^2}{\Lambda^2}\bigg),
\eeqa
such that
\beqa
K_1+K_2&=&\frac{1}{-{\rm
    ad}_{\partial_\mu}{\rm ad}_{\partial_\mu}+M^2}\int_0^1 ds\left[1-\exp\bigg((1-s) \frac{-{\rm ad}_{\partial_\mu}{\rm
      ad}_{\partial_\mu}+M^2}{\Lambda^2}\bigg)\right]\frac{-\frac{i}{2}\,\big\{{\rm ad}_{\partial_\mu},{\rm
  ad}_{\fet{\tau}\cdot\big(\fet{\pi}\times\partial_\mu\fet{\pi}\big)}\big\}-
M^2\fet{\pi}^2}{2 F^2\Lambda^2}\nn
&\times&\exp\bigg(s \frac{-{\rm ad}_{\partial_\mu}{\rm
      ad}_{\partial_\mu}+M^2}{\Lambda^2}\bigg)+{\cal O}(\fet \pi^4).\label{K1PlusK2Expanded}
\eeqa
Further, keeping the terms with up to four pion fields, we obtain
\beqa
-\frac{F^2}{4}{\rm Tr}\,{\rm EOM}\frac{1}{-{\rm ad}_{\partial_\mu}{\rm
      ad}_{\partial_\mu}+M^2}{\rm
    EOM}&=&-\frac{1}{2}\fet{\pi}\cdot\big(-\partial^2+M^2\big)\fet{\pi}+
  \frac{\alpha}{F^2}\fet{\pi}^2\fet{\pi}\cdot\big(-\partial^2+M^2\big)\fet{\pi}\nn
&+&\frac{\fet{\pi}^2}{F^2}\bigg(\partial_\mu\fet{\pi}\cdot\partial_\mu\fet{\pi}+\frac{1}{2}\fet{\pi}\cdot\partial^2\fet{\pi}\bigg)+{\cal
  O}(\fet \pi^6).\label{LRegNoExpExpanded}
\eeqa
For the unregularized pion Lagrangian in Eq.~(\ref{Lpi2Reg}), we have
\beqa
{\cal L}_\pi^{\rm E\, (2)}&=&-F^2
M^2-\frac{1}{2}\fet{\pi}\cdot\big(\partial^2-M^2\big)\fet{\pi}+\frac{\fet{\pi}^2}{8F^2}\Big[-2\partial_\mu\fet{\pi}\cdot\partial_\mu\fet{\pi}-\fet{\pi}\cdot\partial^2\fet{\pi}+(1-8\alpha)\fet{\pi}\cdot(-\partial^2+M^2)\fet{\pi}\Big]+{\cal
  O}(\fet \pi^6).\label{L2UnregExpanded}\quad\quad
\eeqa
Combining Eqs.~(\ref{LRegNoExpExpanded}) and (\ref{L2UnregExpanded}),
we end up with 
\beqa
\label{Temp100}
{\cal L}_\pi^{\rm E\, (2)} - \frac{F^2}{4}{\rm Tr}\,{\rm EOM}\frac{1}{-{\rm ad}_{\partial_\mu}{\rm
      ad}_{\partial_\mu}+M^2}{\rm
    EOM}&=& -F^2
M^2
+\frac{\fet{\pi}^2}{8F^2}\Big[6\,\partial_\mu\fet{\pi}\cdot\partial_\mu\fet{\pi}+3\,\fet{\pi}\cdot\partial^2\fet{\pi}+\fet{\pi}\cdot(-\partial^2+M^2)\fet{\pi}\Big]
\nn
&+&{\cal
  O}(\fet \pi^6).\quad
\eeqa
The exponential operator with the propagator sandwiched between the
EOM terms is given by
\beqa
\label{Temp101}
\frac{F^2}{4}{\rm Tr}\,{\rm EOM}\frac{\exp\Big(\frac{-{\rm ad}_{\partial_\mu}{\rm
      ad}_{\partial_\mu}+M^2}{\Lambda^2}\Big)}{-{\rm ad}_{\partial_\mu}{\rm
      ad}_{\partial_\mu}+M^2}{\rm
    EOM}&=&\frac{1}{2}\fet{\pi}\cdot\big(-\partial^2+M^2\big)e^{\frac{-\partial^2+M^2}{\Lambda^2}}\fet{\pi}
-\frac{\alpha}{F^2}\fet{\pi}^2\fet{\pi}\cdot\big(-\partial^2+M^2\big)e^{\frac{-\partial^2+M^2}{\Lambda^2}}\fet{\pi}
\nn
&
-&\frac{1}{F^2}\bigg(\partial_\mu\fet{\pi}\cdot\partial_\mu\fet{\pi}+\frac{1}{2}\fet{\pi}\cdot\partial^2\fet{\pi}\bigg)\fet{\pi}\cdot
e^{\frac{-\partial^2+M^2}{\Lambda^2}}\fet{\pi}+{\cal
  O}(\fet \pi^6).
\eeqa
Using Eq.~(\ref{K1PlusK2Expanded}), we obtain
\beqa
\label{EOMK1PlusK2EOM}
-\frac{F^2}{4}{\rm Tr} \,{\rm EOM}\big(K_1+K_2\big){\rm
  EOM}&=&
\frac{1}{4 F^2\Lambda^2}\int_0^1
d
s\bigg\{M^2 \fet{\pi}^2\bigg[\bigg(1-e^{(1-s)\frac{-\partial^2+M^2}{\Lambda^2}}\bigg)\fet{\pi}\bigg]
\cdot\bigg[\big(-\partial^2+M^2\big)e^{s\frac{-\partial^2+M^2}{\Lambda^2}}\fet{\pi}\bigg]\nn
&+&\big(\fet{\pi}\times\partial_\mu\fet{\pi}\big)\cdot\bigg[\bigg(e^{s\frac{-\partial^2+M^2}{\Lambda^2}}\big(-\partial^2+M^2\big)\fet{\pi}\bigg)\times
\overleftrightarrow{\partial}_{\!\!\mu}\bigg(1-^{(1-s)\frac{-\partial^2+M^2}{\Lambda^2}}\bigg)\fet{\pi}\bigg]\bigg\}\nn
&+&{\cal
  O}(\fet \pi^6),
\eeqa
where we have used the notation
\beqa
A\cdot\Big[B\times\overleftrightarrow{\partial}_{\!\!\mu} C\Big]&=&A\cdot\Big[B\times\partial_\mu
C\Big]-A\cdot\Big[\big(\partial_\mu B\big)\times C\Big].
\eeqa
The double-cross product in Eq.~(\ref{EOMK1PlusK2EOM}) is generated by the
connection $\Gamma_\mu$. Combining the terms in Eqs.~(\ref{Temp100}),
(\ref{Temp101}) and (\ref{EOMK1PlusK2EOM}), we finally arrive at the 
regularized Lagrangian ${\cal L}_{\pi,\Lambda}^{\rm E \, (2)}$ given in
Eq.~(\ref{regularizedLpi2Expanded}). 

\section{Gradient flow in chiral perturbation theory}
\label{formal_proofs_gradient_flow}

Here, we give a proof that the solution $W$ of the gradient flow
equation (\ref{gradientFlowDefinition}) is an element of $SU(2)$ and
transforms linearly under chiral transformations, i.e.~
$W\to R\,W\,L^\dagger$.

To show  that $W$
is unitary, we consider  the gradient flow equation written in the form 
\beqa
\partial_\tau W W^\dagger&=&-\frac{i}{2} w\big({\rm EOM}(\tau) +{\rm
  EOM}^\dagger(\tau)\big) w^\dagger\,.\label{definition_gradient_flow}  
\eeqa
Equation (\ref{definition_gradient_flow}) is identical to
Eq.~(\ref{gradientFlowDefinition}) if the matrix $W$ 
is unitary. If we take
Eq.~(\ref{definition_gradient_flow}) as a definition of the gradient flow
equation, then the unitarity of $W$ follows immediately by taking hermitian
conjugate of this equation, yielding  
\beqa
W\partial_\tau W^\dagger&=&\frac{i}{2}w\big({\rm EOM}(\tau) +{\rm
  EOM}^\dagger(\tau)\big) w^\dagger\,.\label{definition_gradient_flow_hc}  
\eeqa
Adding Eqs.~(\ref{definition_gradient_flow}) and
(\ref{definition_gradient_flow_hc}), we obtain
\beqa 
\partial_\tau\big(W\,W^\dagger\big)=\partial W\,W^\dagger +
W\partial_\tau W^\dagger&=&0\quad\Rightarrow\quad W\,W^\dagger\,=\,
{\rm const}\,.
\eeqa

To show that $\det W=1$, we use the Jacobi formula for a derivative of a
determinant:
\beqa
\partial_\tau \det W&=& \det W\,{\rm Tr}\big(W^{-1} \partial_\tau W\big)\,.\label{Jacobi_Formula}
\eeqa
Exploiting the unitarity of $W$ and using the gradient flow
Eq.~(\ref{definition_gradient_flow_hc}), we obtain from Eq.~(\ref{Jacobi_Formula})
\beqa
\partial_\tau \det W&=& -\frac{i}{2}\det W\,{\rm Tr}\Big[W^\dagger
w\big({\rm EOM}(\tau)+{\rm EOM}^\dagger(\tau)\big)w\Big]\nonumber\\
&=& -\frac{i}{2}\det W\,{\rm Tr}\big[{\rm EOM}(\tau)+{\rm EOM}^\dagger(\tau)\big]\,=\,0\,,\label{Derivative_of_det}
\eeqa
where in the last step of Eq.~(\ref{Derivative_of_det}) we used that
the ${\rm EOM}(\tau)$ operator is
traceless (by
construction). Since $\det W$ does not depend on $\tau$ and $\det
W(\tau=0)=\det U=1$, it follows that $\det W=1$. This implies that $W$ is
an element of the SU$(2)$ group for all values of $\tau$.

To
prove the chiral transformation properties of $W$, we rewrite the EOM$(\tau)$
operator via
\beqa
-i\, w\,{\rm EOM}(\tau)\,w&=&\frac{1}{2}\bigg[(\nabla_\mu\nabla_\mu W)W^\dagger-W(\nabla_\mu\nabla_\mu
W)^\dagger+\chi W^\dagger - W\chi^\dagger  - \frac{1}{2}{\rm Tr}\big(\chi W^\dagger - W\chi^\dagger \big) \bigg]\,W\,.
\eeqa
Here, we used the definition of the EOM in Eqs.~(\ref{EOM_Second_Step}) and
(\ref{EOM_Final_Step}). Thus, the gradient flow equation (\ref{gradientFlowDefinition}) is
rewritten to
\beqa
\partial_\tau W&=&\frac{1}{2}\bigg[(\nabla_\mu\nabla_\mu W)W^\dagger-W(\nabla_\mu\nabla_\mu
W)^\dagger+\chi W^\dagger - W\chi^\dagger  - \frac{1}{2}{\rm Tr}\big(\chi W^\dagger - W\chi^\dagger \big) \bigg]\,W\,.\label{gradient_flow_in_terms_of_W}
\eeqa
Equation (\ref{gradient_flow_in_terms_of_W}) is more convenient for
studying the chiral transformation properties, since everything here is
expressed in terms of $W$. Next, we perform the small-$\tau$
expansion 
\beqa
W &=&\sum_{n=0}^\infty \tau^n W^{(n)}\,,
\eeqa
where $W^{(n)}$ does not depend on
$\tau$. Eq.~(\ref{gradient_flow_in_terms_of_W}) then becomes
\beqa
\sum_{n=0}^\infty(n+1)\tau^n
W^{(n+1)}&=&\frac{1}{2}\chi+\sum_{n=0}^\infty\tau^n \frac{1}{2}\bigg[\nabla_\mu\nabla_\mu
W^{(n)} -\sum_{n_1+n_2+n_3=n}W^{(n_1)}(\nabla_\mu\nabla_\mu
W^{(n_2)})^\dagger W^{(n_3)}\nn
&-& \sum_{n_1+n_2=n}\bigg(W^{(n_1)}\chi^\dagger W^{(n_2)}+\frac{1}{2}{\rm
  Tr}\big[\chi  W^{(n_1)\dagger} - W^{(n_1)}\chi^\dagger \big] W^{(n_2)}\bigg)\bigg]\,.\label{gradient_flow_expansion_in_tau}
\eeqa
Eq.~(\ref{gradient_flow_expansion_in_tau}) provides a recursion
relation, which allows one to verify the transformation properties of
the $W^{(n)}$ operators. We start the induction with $n=0$ and obtain
\beqa
W^{(1)}&=&\frac{1}{2}\chi+\frac{1}{2}\bigg[\nabla_\mu\nabla_\mu
W^{(0)}-W^{(0)}(\nabla_\mu\nabla_\mu W^{(0)})^\dagger
W^{(0)}-W^{(0)}\chi^\dagger W^{(0)}-\frac{1}{2}{\rm
  Tr}\big(\chi  W^{(0)\dagger} - W^{(0)}\chi^\dagger \big) W^{(0)}\bigg]\nn
&=&\frac{1}{2}\chi+\frac{1}{2}\bigg[\nabla_\mu\nabla_\mu
U-U(\nabla_\mu\nabla_\mu U)^\dagger
U-U\chi^\dagger U-\frac{1}{2}{\rm
  Tr}\big[\chi  U^{\dagger} -U\chi^\dagger \big] U\bigg]\,.
\eeqa
Under chiral rotations, the field $U$, its covariant derivative
and the source $\chi$ transform via
\beqa
\chi&\to&R \,\chi\,L^\dagger, \quad\quad  U\,\to\,R\,U\,L^\dagger,
\quad\quad \nabla_\mu U\,\to\,R\,\nabla_\mu U L^\dagger\,.
\eeqa
Thus, we see that $W^{(1)}$ transforms as
\beqa
W^{(1)}&\to&R\,W^{(1)}\,L^\dagger\,. \label{chiralTransfW}
\eeqa
Notice that covariant derivatives $\nabla_\mu$ do not depend on $W$:
\beqa
\nabla_\mu W^{(1)}=\partial_\mu W^{(1)} - i\, r_\mu W^{(1)} +i\, W^{(1)} l_\mu.
\eeqa
Since $W^{(1)}$ has the same behaviour under chiral transformations as
$U$, the covariant derivative of $W^{(1)}$ transforms as
\beqa
\nabla_\mu W^{(1)}&\to&R\,\nabla_\mu W^{(1)}\,L^\dagger\,.
\eeqa
Suppose now that $W^{(j)}$ transforms as
\beqa
W^{(j)}&\to&R\,W^{(j)}\,L^\dagger\,,
\eeqa
for $j\leq n$. 
Then, Eq.~(\ref{gradient_flow_expansion_in_tau}) fixes the
transformation behaviour of $W^{(n+1)}$:
\beqa
W^{(n+1)}&=&\frac{1}{n+1}\bigg[\frac{1}{2}\nabla_\mu\nabla_\mu
W^{(n)} -\sum_{n_1+n_2+n_3=n}W^{(n_1)}(\nabla_\mu\nabla_\mu
W^{(n_2)})^\dagger W^{(n_3)}\nn
&-& \sum_{n_1+n_2=n}\bigg(W^{(n_1)}\chi^\dagger W^{(n_2)}+\frac{1}{2}{\rm
  Tr}\big[\chi  W^{(n_1)\dagger} - W^{(n_1)}\chi^\dagger \big] W^{(n_2)}\bigg)\bigg]\,.\label{gradient_flow_recursion_rel}
\eeqa
Since on the right-hand side of
Eq.~(\ref{gradient_flow_recursion_rel}) all superscripts of $W$ are
smaller or equal $n$, we conclude that
\beqa
W^{(n+1)}&\to&R\,W^{(n+1)}\,L^\dagger.
\eeqa
This proves the desired relationship 
\beqa
W&\to&R\,W\,L^\dagger.
\eeqa


\end{document}